\documentclass[sigconf,authorversion,screen]{acmart}

\AtBeginDocument{%
  }

\usepackage{subcaption}
\usepackage{dblfloatfix}
\usepackage{bm}

\usepackage{multirow}
\usepackage{makecell}
\usepackage{array}

\usepackage[capitalise]{cleveref}
\newcommand{\etal}{\textit{et al.}}
\newcommand{\eg}{\textit{e.g.}}

\usepackage{soul}

\definecolor{darkgreen}{RGB}{0,128,0}

\newcommand{\vect}[1]{\mathbf{#1}}
\newcommand{\mat}[1]{\mathbf{#1}} %

\newcommand{\vz}{\vect{z}}
\newcommand{\vy}{\vect{y}}
\renewcommand{\vv}{\vect{v}}
\newcommand{\vepsilon}{\boldsymbol{\epsilon}}
\newcommand{\vc}{\vect{c}}
\newcommand{\mM}{\mat{M}}
\newcommand{\mA}{\mat{A}}
\newcommand{\mB}{\mat{B}}
\newcommand{\mW}{\mat{W}}
\newcommand{\mI}{\mat{I}}
\newcommand{\mTheta}{\bm{\Theta}}
\newcommand{\vg}{\vect{g}}

\usepackage[many]{tcolorbox}
\usepackage[table]{xcolor}
\usepackage{fontawesome}

\newcolumntype{Y}[1]{>{\centering\arraybackslash}p{#1}}

\colorlet{mygreen}{green!60!black}
\colorlet{myred}{red}

\begin{document}

\title{DiV-INR: Extreme Low-Bitrate Diffusion Video Compression with INR Conditioning}

\author{Eren \c{C}etin}
\affiliation{%
  \institution{ETH Z\"urich}
  \city{Z\"urich}
  \country{Switzerland}
}
\author{Lucas Relic}
\affiliation{%
  \institution{ETH Z\"urich}
  \city{Z\"urich}
  \country{Switzerland}
}
\author{Yuanyi Xue}
\affiliation{%
  \institution{Disney Entertainment and ESPN Product \& Technology}
  \city{San Francisco}
  \country{USA}
}
\author{Markus Gross}
\affiliation{%
  \institution{ETH Z\"urich}
  \city{Z\"urich}
  \country{Switzerland}
}
\author{Christopher Schroers}
\affiliation{%
  \institution{DisneyResearch\textbar Studios}
  \city{Z\"urich}
  \country{Switzerland}
}
\author{Roberto Azevedo}
\affiliation{%
  \institution{DisneyResearch\textbar Studios}
  \city{Z\"urich}
  \country{Switzerland}
}

\begin{teaserfigure}
  \centering
  \includegraphics[width=\textwidth]{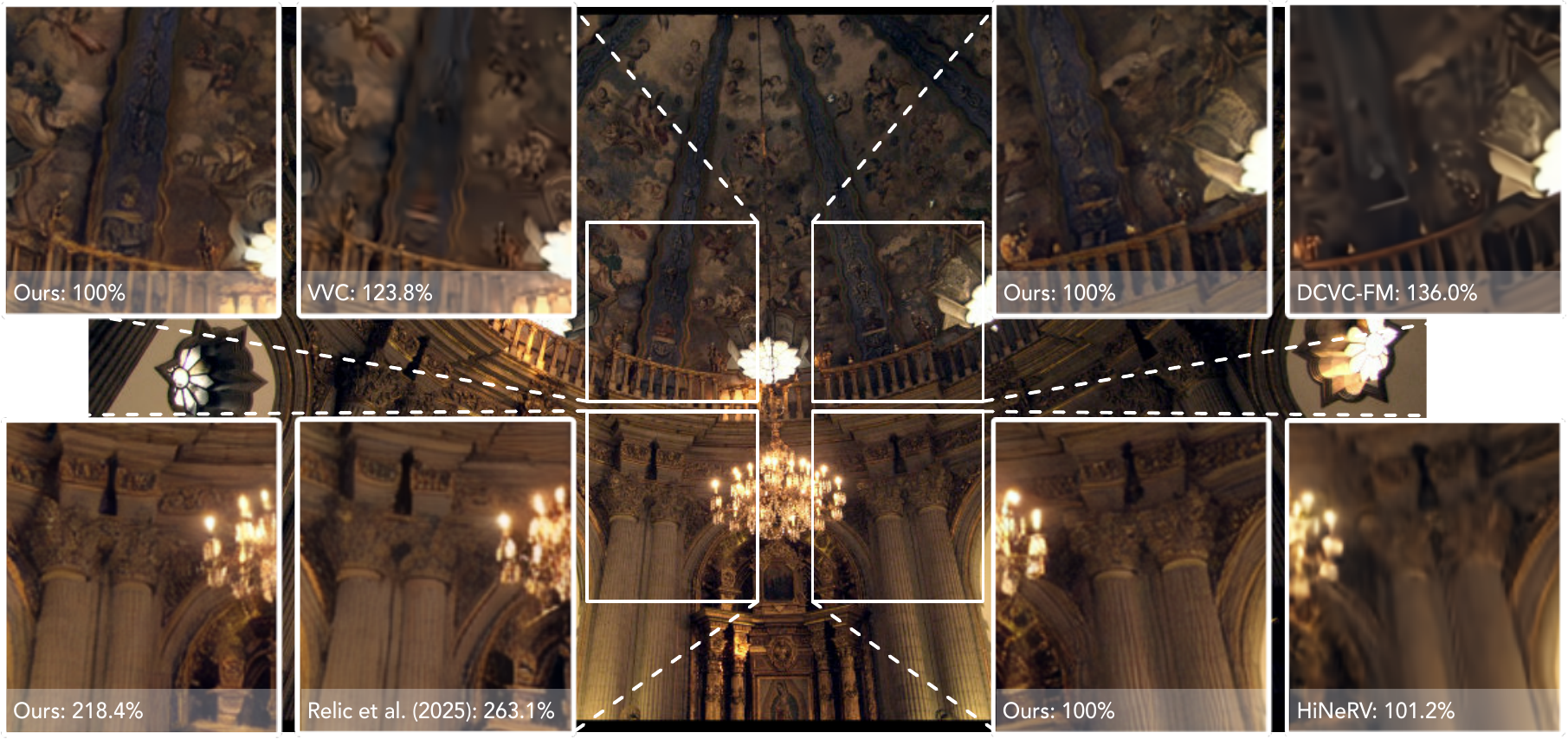}
  \caption{\textbf{Illustrative comparison of our approach with baselines.}
    Our video compression method maintains pleasing details and high fidelity even at
    extremely low bitrates, whereas traditional codecs~(VVC~\cite{vvcoverview}), deep learned
    methods~(DCVC-FM~\cite{li2024neural}), diffusion-based compression~(Relic~\etal~\cite{relic2025spatiotemporal}),
    and INR approaches~(HiNeRV~\cite{kwanHiNeRVVideoCompression2024}) introduce blur or noise
    at comparable rates. For pairwise comparison, crops show our method at two different rates;
    percentages reflect bitrate relative to our lowest rate model (0.0186\,bpp).
    Best viewed digitally.}
  \label{fig:teaser}
\end{teaserfigure}

\begin{abstract}
  We present a perceptually-driven video compression framework integrating implicit neural representations~(INRs) and pre-trained video diffusion models to address the extremely low bitrate regime~($<0.05$~bpp).
  Our approach exploits the complementary strengths of INRs, which provide a compact video representation, and diffusion models, which offer rich generative priors learned from large-scale datasets.
  The INR-based conditioning replaces traditional intra-coded key\-frames with bit-efficient neural representations trained to estimate latent features and guide the diffusion process.
  Our joint optimization of INR weights and parameter-efficient adapters for diffusion models allows the model to learn reliable conditioning signals while encoding video-specific information with minimal parameter overhead.
  Our experiments on UVG, MCL-JCV, and JVET Class-B benchmarks demonstrate substantial improvements in perceptual metrics~(LPIPS, DISTS, and FID) at extremely low bitrates, including improvements on BD-LPIPS up to $0.214$ and BD-FID up to $91.14$ relative to HEVC, while also outperforming VVC and previous strong state-of-the-art neural and INR-only video codecs.
  Moreover, our analysis shows that INR-conditioned diffusion-based video compression first composes the scene layout and object identities before refining textural accuracy, exposing the semantic-to-visual hierarchy that enables perceptually faithful compression at extremely low bitrates.
\end{abstract}



\maketitle

\section{Introduction}
\label{sec:introduction}

Video compression at extremely low bitrates poses a fundamental challenge for both conventional codecs (e.g., AV1~\cite{han2021technicaloverviewav1}, VVC~\cite{vvcoverview}) and recent neural video compression~(NVC) methods~\cite{li2021Deep,zhenghao2023neural,li2024neural}.
As the bitrate falls below \(\sim\)0.05 bits per pixel, these methods struggle to preserve high-frequency details, leading to severe blurring, blocking, or banding artifacts.
To tackle this problem, generative video compression methods leveraging video diffusion models~\cite{relic2025spatiotemporal,li2024extremevideocompressionpretrained} have emerged as a promising alternative.

By employing robust priors to generate missing details, diffusion-based codecs excel at synthesizing realistic, high-fidelity reconstructions even at extreme compression ratios.
To preserve fidelity to the original source video, current video diffusion-based codecs use heavily compressed keyframes~\cite{li2024extremevideocompressionpretrained} or explicit optical flow~\cite{relic2025spatiotemporal} to guide the generative process.
However, this localized and sparse conditioning signal often provides inadequate temporal guidance, which can lead the generative model to create incorrect structures or lose temporal consistency far from a keyframe.

Implicit neural representations~(INRs)~\cite{dupont2021coin, sitzmannImplicitNeuralRepresentations2020} offer a fundamentally different approach to video representation.
By training a continuous coordinate network to fit to a specific video, INRs create highly compact representations that capture the global spatiotemporal features of an entire sequence.
While standalone INRs may struggle to render fine, high-frequency details, especially at very low bitrates, their ability to efficiently encode the global sequence features makes them an excellent candidate to guide the diffusion decoding process.

Motivated by the complementary strengths of INRs and diffusion models, we propose a novel video compression framework, DiV-INR, that integrates the compact video representation of INRs with the rich generative capabilities of diffusion models.
Specifically, we use an INR encoded at the sender to produce features to condition a video diffusion decoder, guiding it to produce a detailed and high-fidelity reconstruction.
To enhance the effectiveness of this conditioning, we optimize the INR directly within the loop of the diffusion model, enabling it to learn a representation that is explicitly tailored to the generative decoder.
Furthermore, since this conditioning is inherently specific to each instance, we employ parameter-efficient fine-tuning~(PEFT)~\cite{koohpayeganiNOLACompressingLoRA2024} to adapt the diffusion model to the unique signal of the INR with minimal parameter overhead~(<0.1\% of the base model).

Extensive experiments on standard benchmarks (UVG~\cite{uvg_dataset}, MCL-JCV~\cite{wang2016mcl}, and JVET Class-B~\cite{hevc}) demonstrate that our proposed framework yields significant perceptual gains at extremely low bitrates when compare to the state of the art, while remaining practical on standard consumer hardware.

\section{Related Work}
\label{sec:related}

\paragraph{Implicit neural video representations~(INRs).}
INRs provide a highly compact alternative for signal representation by encoding an image or video directly into the weights of a coordinate-based neural network~\cite{dupont2021coin}.
This paradigm fundamentally reframes video coding from a explicit data compression problem to a neural network model compression problem.
This approach was pioneered for video by NeRV~\cite{chenNeRVNeuralRepresentations2021}, which overfits a network mapping temporal frame indices to RGB frames, achieving performance comparable to HEVC with $25\times$ faster encoding.
Subsequent works focused on improving network parameterization, for instance, by disentangling spatial and temporal contexts~\cite{li2022enervexpediteneuralvideo}, explicitly modeling inter-frame dynamics~\cite{zhao2023dnervmodelinginherentdynamics}, or utilizing a combination of frame and GOP-level tokens~\cite{saethreCombiningFrameGOP2024}.
To overcome the representation bottleneck of purely coordinate-based inputs, other methods jointly optimize learnable feature grids which are passed to the decoding network~\cite{chen2023hnervhybridneuralrepresentation, kwanHiNeRVVideoCompression2024}.
Beyond improving network parameterization, these approaches maximize compression efficiency through a combination of architectural upgrades (\eg, ConvNeXt blocks~\cite{chen2023hnervhybridneuralrepresentation} or depthwise convolutions~\cite{kwanHiNeRVVideoCompression2024}), post-training hierarchical pruning, or end-to-end rate-distortion training under explicit entropy constraints~\cite{gomesVideoCompressionEntropyConstrained2023a}.

However, these methods extract final pixel values directly from the INR representation and optimize for pixel-domain fidelity.
While this yields compact bitstreams, perceptual quality remains limited
at extreme bitrates where the INR lacks the capacity to synthesize realistic details.
We instead deploy a lightweight INR strictly as a conditioning signal for pre-trained diffusion models, allowing the framework to generate realistic results even at extremely low bitrates.

\paragraph{Generative diffusion compression.}
First employed in the image domain~\cite{relic2024lossyimagecompression, yang2023Lossy, xia2024DiffPC}, diffusion-based image codecs regenerate the source image at the receiver using semantic~\cite{xia2024DiffPC} or structural features~\cite{yang2023Lossy,relic2024lossyimagecompression} extracted from the input.
Current methods in the video domain follow a similar conditional generation paradigm; however, they typically restrict their conditioning signals to localized pixel-level frames or explicit motion representations.
For instance, Li~\etal~\cite{li2024extremevideocompressionpretrained} transmit independently compressed keyframes and generate the video at the receiver using this context.
When the quality of generated frames falls below a predefined threshold, new keyframes are sent and the process continues.
Relic~\etal~\cite{relic2025spatiotemporal} framed compression as an interpolation problem, sending the first and last frame in a group of pictures~(GoP) and synthesizing the intermediate frames in between.
To preserve fidelity, they transmit explicit optical flow to warp the keyframes into intermediate predictions, which serves as additional generative guidance.
Although bit-efficient, this reliance on warping inherently fails when intermediate frames contain new or occluded content not visible in the keyframes.

Gao~\etal~\cite{gao2025givicgenerativeimplicitvideo} introduced GiViC, a generative implicit video compression framework that embedded a diffusion process within an INR framework.
They augment the INR decoder with a diffusive sampling process across cascaded spatiotemporal pyramids to capture long-range dependencies across the sequence; however, the underlying denoising model still relies on a relatively simple, MLP-based architecture trained from scratch.
Consequently, it lacks the representational capacity and rich visual priors inherent to large-scale foundation models, limiting its generative capabilities at extremely low bitrates.

Unlike the above approaches that utilize sub-optimal hand-craft\-ed conditioning signals or small denoisers trained from scratch, we leverage INRs to produce per-instance optimized conditioning features and employ parameter-efficient adaptation of a large pre-trained foundation diffusion models~\cite{wan2025wanopenadvancedlargescale,blattmann2023stablevideodiffusionscaling,yangCogVideoXTexttoVideoDiffusion2024}, allowing the strong generative prior to fully exploit the dense INR features in an efficient manner.

\begin{figure*}[t]
  \centering
  \includegraphics[width=\linewidth]{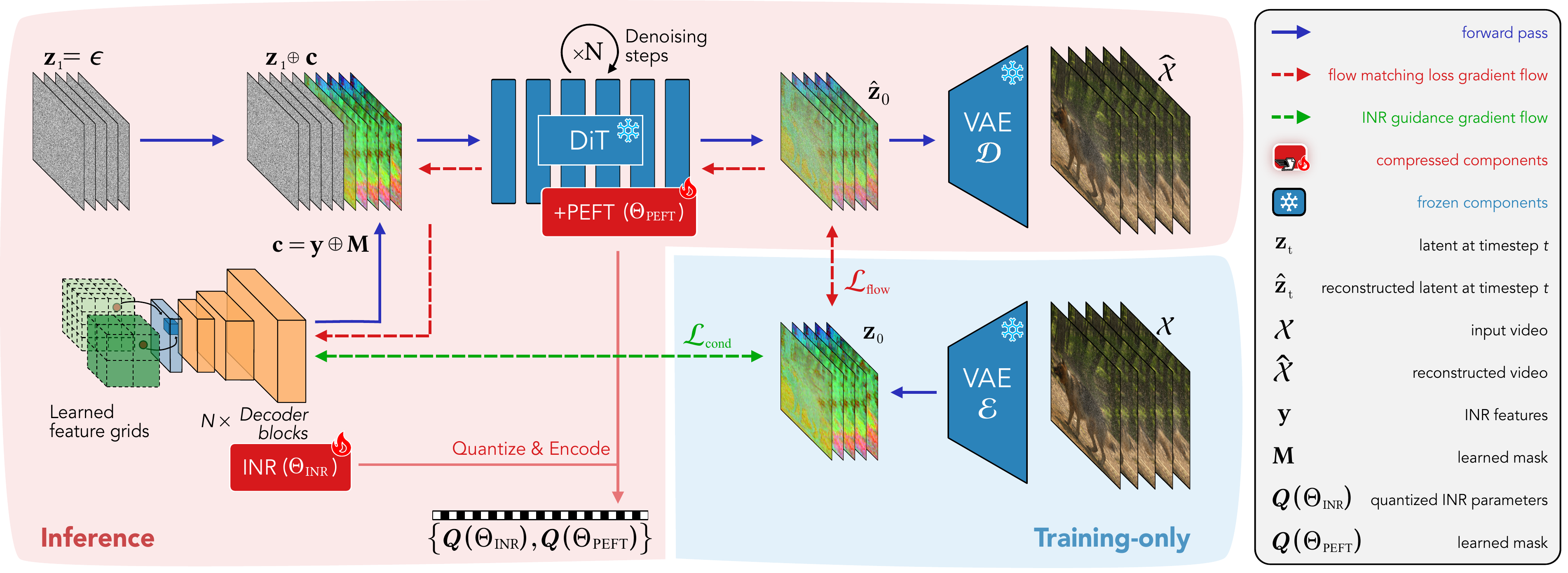}
  \caption{\textbf{Architecture of our proposed framework.} An INR-driven conditioning module and
    a parameter-efficient adapter tailor the video diffusion model. The INR produces conditioning
    signals and masks, while the adapter specializes the model to the video content and motion
    characteristics. Both sets of weights are quantized and arithmetically encoded for compression.}
  \label{fig:overview}
\end{figure*}

\section{Method}
\label{sec:method}

Let $\mathcal{X} \in \mathbb{R}^{T \times H\times W \times 3}$ be a sequence of $T$ consecutive frames with spatial resolution $H \times W$.
Our goal is to compress $\mathcal{X}$ into a compact representation that enables high-fidelity reconstruction through a pre-trained video diffusion model.

As shown in \cref{fig:overview}, our video compression framework produces a compressed representation consisting of: i)~$\mTheta_{\text{INR}}$, the parameters of an INR that generates conditioning signals; and ii)~$\mTheta_{\text{PEFT}}$, the adapter weights of a pre-trained video diffusion model for an instance-specific model adaptation.

The reconstructed video $\hat{\mathcal{X}} \in \mathbb{R}^{T \times H\times W \times 3} $ is obtained as:
\begin{equation}
  \hat{\mathcal{X}} =
  \mathcal{D}\left(\text{DiT}\left(\vz_1,\,\vg\left(f;\,\mTheta_{\text{INR}}\right);\,\mTheta_{\text{PEFT}}\right)\right) \\
\end{equation}
where $\mathcal{D}$ denotes the VAE decoder, $\text{DiT}$ denotes the multistep denoising process performed by the diffusion transformer on sample $\vz_1\sim\mathcal{N}(\mathbf{0},\mI)$ from Gaussian prior distribution conditioned on features encoded by INR $\vg$, sampled for the frame/s of interest, $f$.

The video diffusion model operates in a latent space obtained through a 3D causal VAE encoder, $\mathcal{E}$~\cite{wan2025wanopenadvancedlargescale}.
The encoder compresses the input video with
 spatio-temporal downsampling, producing latent representations $\vz_0 = \mathcal{E}\left(\mathcal{X}\right) \in \mathbb{R}^{T' \times H' \times W'\times C}$, where $C$ is the number of latent channels, and $T', (H', W')$ denote the downsampled temporal and spatial dimensions, respectively.
Specifically, we adopt Wan2.1~\cite{wan2025wanopenadvancedlargescale} as our video diffusion model.

\subsection{INR-Based Adaptive Conditioning}
\label{subsec:inr_conditioning}

One of our core contributions is a conditioning mechanism that replaces traditional intra-coded keyframes with compact INRs optimized to guide the diffusion process.
Unlike conventional approaches that use fixed conditioning signals~(\eg, compressed key\-frames from standard codecs), our method learns to produce conditioning signals specifically tailored for generative reconstruction.
This approach jointly trains neural representations and parameter-efficient diffusion adapters,
enabling video-specific adaptation with minimal parameter overhead while maintaining strong
generative priors.

\paragraph{INR architecture and conditioning.}
The INR is parameterized as $\vg: \mathbb{R} \to \mathbb{R}^{C \times H'
\times W'} \times [0,1]^{C_m \times H' \times W'}$ that maps normalized temporal coordinates $f \in
[0,1]$ to latent-space conditioning:
\begin{equation}
  \vg(f;\mTheta_{\text{INR}}) = (\vy_f,\;\mM_f)
\end{equation}
where $\vy_f \in \mathbb{R}^{C \times H' \times W'}$ is the predicted conditioning signal
for latent frame $f$,
and $\mM_f \in [0,1]^{C_m \times H' \times W'}$ is an adaptive temporal mask with
$C_m{=}4$ channels quantifying the ``confidence'' in the conditioning signal present
in each conditional latent frame, $\vy_f$.
The network consists of a learned feature grid followed by a convolutional decoder similar to
HiNeRV~\cite{kwanHiNeRVVideoCompression2024}. In contrast to HiNeRV, our INR encodes a latent
representation that is suitable for conditioning the diffusion model with a learned mask and
conditioning information, rather than generating RGB frames.

For each GoP, the INR generates the conditioning information for the latent frames of the GoP, $\vy \in \mathbb{R}^{C
\times T' \times H' \times W'}$ and mask $\mM \in [0,1]^{C_m \times T' \times H' \times W'}$ by evaluating $\vg_{\mTheta_{\text{INR}}}$ at uniformly spaced temporal coordinates. The mask provides adaptive weighing, indicating regions where INR predictions are reliable versus areas requiring greater reliance on the diffusion prior. Our bitstream contains \emph{no} traditional I-frames; the INR replaces keyframe conditioning entirely, providing continuous temporal guidance.
The final conditioning
signal $\vc = \text{concat}(\vy,\;\mM)$ is concatenated with noisy latents during diffusion,
providing rich temporal guidance throughout the denoising process.

\subsection{Parameter-efficient adaptation}
\begin{figure}
  \centering
  \includegraphics[width=\linewidth]{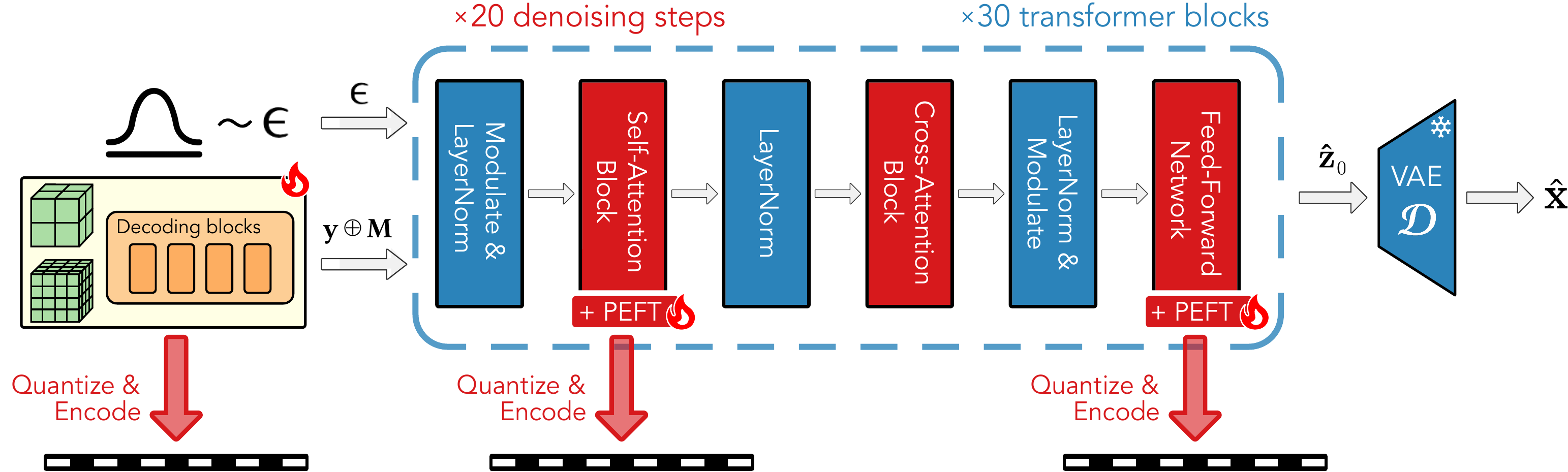}
  \caption{
    \textbf{Adapter placement with INR conditioning.} The INR branch yields latent
    conditioning signals and masks, while the diffusion transformer is augmented with
    NOLA adapters in output transformations of self-attention layers and feed-forward layers
    to specialize for video-specific motion without inflating bitrate.
  }
  \label{fig:nola}
\end{figure}

While the INR provides temporal conditioning, video-specific adaptation of the diffusion model itself is necessary to align the generative prior with target content characteristics.
For this purpose, we employ NOLA~\cite{koohpayeganiNOLACompressingLoRA2024} adapters that enable fine-tuning of the pre-trained diffusion backbone without prohibitive parameter overhead.
NOLA extends Low-Rank Adaptation~(LoRA)~\cite{huLoRALowRankAdaptation2021} by reparameterizing weight updates as linear combinations of fixed pseudo-random bases:
\begin{align}
  \mW = \mW_0 +
  \left( \sum_{i=1}^{b} {\beta_i} \mB^{(i)} \right)
  \left( \sum_{i=1}^{b} {\alpha_i} \mA^{(i)} \right)
\end{align}
where $\mW_0\in\mathbb{R}^{m\times n}$ is the pre-trained weight matrix, $\left\{\mA^{(i)},
  \mB^{(i)}\right\}_{i=1}^b \sim \mathcal{N}(\mathbf{0},s\mathbf{I})$ are the pseudo-random bases and
$s=0.25$ is a scaling factor set to tune the impact of adapters.
The basis matrices are drawn once from a known seed and then frozen.
Only scalar coefficients $\{{\alpha_i},\,{\beta_i}\}$ are trained for $b$ basis matrices~(a total of $2b$ parameters) per low-rank mapping matrix. We inject NOLA adapters with rank $r{=}64$, into $30$ DiT blocks, targeting self-attention output projections, feed-forward layers and the final output
head of the diffusion transformer, as depicted in \cref{fig:nola}.
Using a basis set of $500$ random matrices, the PEFT adapter
requires only $91{,}000$ trainable parameters, achieving ${\sim}25\times$ parameter reduction per layer compared to LoRA~\cite{huLoRALowRankAdaptation2021} with rank $r{=}16$ while also eliminating the dependence between the diffusion model's architectural choices and the bitrate, since the bitrate
is no longer a function of the rank and the hidden dimension of the transformer blocks.

\subsection{Pruning and quantization}

To achieve extreme compression rates, we apply unstructured model compression to both the INR and
NOLA adapter weights, similar to HiNeRV~\cite{kwanHiNeRVVideoCompression2024}.
These techniques exploit the redundancy in neural network parameters while maintaining
reconstruction quality.

\paragraph{Adaptive magnitude pruning.}
We apply unstructured pruning exclusively to the INR decoder, removing $15\%$ of parameters using
adaptive magnitude-based scoring~\cite{kwanHiNeRVVideoCompression2024}. Unlike vanilla magnitude
pruning, the adaptive criterion accounts for layer width to prevent excessive removal from narrow
layers such as the initial and final layers of INR decoder:
\begin{equation}
  \text{score}(\theta_i) = \frac{|\theta_i|}{\sqrt{P}}
\end{equation}
where $P$ is the number of parameters in the layer. Weights with scores below the $15$-th
percentile are pruned and remain zero throughout subsequent training. NOLA coefficients are
not pruned due to their inherent compactness.

\paragraph{Quantization-aware training.}
Following pruning, we employ Qu\-ant-Noise~\cite{fanTrainingQuantizationNoise2021} to prepare both
INR and NOLA weights for low-bit quantization. During each forward pass, a fraction $\rho{=}0.9$ of
weight tensors are randomly replaced with their quantized counterparts. We apply $6$-bit uniform
quantization to INR weights and NOLA coefficients at inference, storing per-tensor scale and
zero-point values alongside quantized integers.

\subsection{Training}

Our training procedure jointly optimizes the INR conditioning network and NOLA adapters to compress
video content while maintaining high perceptual quality through the diffusion prior.

\paragraph{Optimization objective.}
\label{par:optimization_objective}
We employ a dual loss that balances diffusion model fidelity with conditioning quality:
\begin{equation}
\begin{split}
  \mathcal{L}_{\text{flow}}  & = \mathbb{E}_{\vz,\vepsilon,t}\big[\|\text{DiT}(\vz_t,t,\vy) -
  \vv^{\star}(\vz_t,t)\|^2\big] \\
  \mathcal{L}_{\text{cond}}  & =\|\vy-\vz_0\|_2^2 \\
  \mathcal{L}_{\text{total}} & = (1 - \lambda_{\text{cond}}) \cdot \mathcal{L}_{\text{flow}} +
  \lambda_{\text{cond}} \cdot \mathcal{L}_{\text{cond}} 
\end{split}
\label{eq:loss-total}
\end{equation}

where $\mathcal{L}_{\text{flow}}$ is the flow-matching loss~\cite{lipman2023fm} drawing timestep
$t$ from log-normal variance-preserving schedule that forms a partially noised latent
$\vz_t=(1 - t)\vz_0+t\vepsilon$ by mixing the clean VAE latent $\vz_0\sim p_{\text{data}}$ with
noise $\vepsilon\sim\mathcal{N}(\mathbf{0},\mI)$, and regresses the predicted velocity towards
the ground truth velocity $\vv^{\star}=\partial \vz_t / \partial t = \vepsilon - \vz_0$, while
$\mathcal{L}_{\text{cond}}$ supervises INR reconstructions. We cosine anneal
$\lambda_{\text{cond}}$ to let the INR dominate early iterations before $\mathcal{L}_{\text{flow}}$
becomes the main supervision.

\paragraph{Three-stage training curriculum.}
We adopt a progressive compression strategy: \textbf{(1)~Dense training} jointly optimizes the INR
($\text{lr}{=}2e{-}3$) and NOLA adapters ($\text{lr}{=}2e{-}4$) with AdamW and cosine decay,
training from scratch on each video GoP for $300$ epochs; \textbf{(2)~Pruning-aware finetuning}
removes $15\%$ of the INR decoder weights and continues optimization to recover performance for $120$
epochs; \textbf{(3)~Quan\-ti\-za\-tion-aware fine\-tun\-ing} enables Quant-Noise with $\rho{=}0.9$ and
reduces the INR learning rate by $10{\times}$ for stable convergence under quantization noise for $60$
epochs. The Wan2.1 backbone remains frozen throughout all stages.

\section{Experiments}
\label{sec:results}

\begin{table*}[b]
  \centering
  \caption{\textbf{BD-metric deltas on UVG, JVET-B, and MCL-JCV.}
    Each entry reports the BD-metric difference between the listed codec and DiV-INR.
    For PSNR~(dB) positive values denote advantage, while for the other perceptual metrics it
    indicates perceptual disadvantages.} \label{tab:bd-rate-gains}
  \renewcommand{\arraystretch}{1.1} \setlength{\tabcolsep}{4pt} \small \resizebox{\textwidth}{!}{%
    \begin{tabular}{@{}r|cccc|cccc|cccc@{}}
      \specialrule{0.8pt}{0pt}{2pt}
      \textbf{BD-Metric}                           &
      \multicolumn{4}{c|}{\textbf{UVG}}            &
      \multicolumn{4}{c|}{\textbf{JVET-B}}         &
      \multicolumn{4}{c}{\textbf{MCL-JCV}}                                                                                                                                                                                                                                                       \\[-1pt]
      \midrule
      \textbf{Codec}                               &
      \textbf{PSNR} $\uparrow$                     & \textbf{LPIPS} $\downarrow$ & \textbf{DISTS} $\downarrow$ & \textbf{FID} $\downarrow$ &
      \textbf{PSNR} $\uparrow$                     & \textbf{LPIPS} $\downarrow$ & \textbf{DISTS} $\downarrow$ & \textbf{FID} $\downarrow$ &
      \textbf{PSNR} $\uparrow$                     & \textbf{LPIPS} $\downarrow$ & \textbf{DISTS} $\downarrow$ & \textbf{FID} $\downarrow$                                                                                                                                                       \\
      \specialrule{0.5pt}{2pt}{2pt}
      H.265/HEVC~\cite{hevc}                       & +1.02                       & +0.195                      & +0.122                    & +72.89        & +2.52          & +0.214         & +0.105         & +91.14        & +6.13          & +0.069         & +0.053         & +33.86        \\
      H.266/VVC~\cite{vvcoverview}                 & +3.13                       & +0.131                      & +0.104                    & +47.95        & +4.37          & +0.160         & +0.103         & +83.46        & +7.31          & +0.052         & +0.063         & +30.18        \\
      DCVC-FM~\cite{li2024neural}                  & \textbf{+3.46}              & +0.156                      & +0.132                    & +71.81        & \textbf{+4.73} & +0.175         & +0.126         & +106.57       & \textbf{+8.41} & +0.049         & +0.065         & +36.51        \\
      Relic~\etal~\cite{relic2025spatiotemporal}   & -3.34                       & +0.064                      & +0.032                    & +7.57         & -2.24          & +0.055         & +0.021         & +13.23        & -1.20          & +0.013         & +0.015         & +5.24         \\
      HiNeRV~\cite{kwanHiNeRVVideoCompression2024} & +3.05                       & +0.108                      & +0.106                    & +40.84        & +2.63          & +0.165         & +0.120         & +71.90        & +5.33          & +0.082         & +0.094         & +35.58        \\
      \specialrule{0.5pt}{2pt}{2pt}
      \rowcolor[HTML]{FAFAFA}
      DiV-INR (ours)                               & 0.00                        & \textbf{0.000}              & \textbf{0.000}            & \textbf{0.00} & 0.00           & \textbf{0.000} & \textbf{0.000} & \textbf{0.00} & 0.00           & \textbf{0.000} & \textbf{0.000} & \textbf{0.00} \\
      \specialrule{0.8pt}{2pt}{0pt}
    \end{tabular}%
  }
\end{table*}

\begin{figure*}[b]
  \centering
  \includegraphics[width=\linewidth]{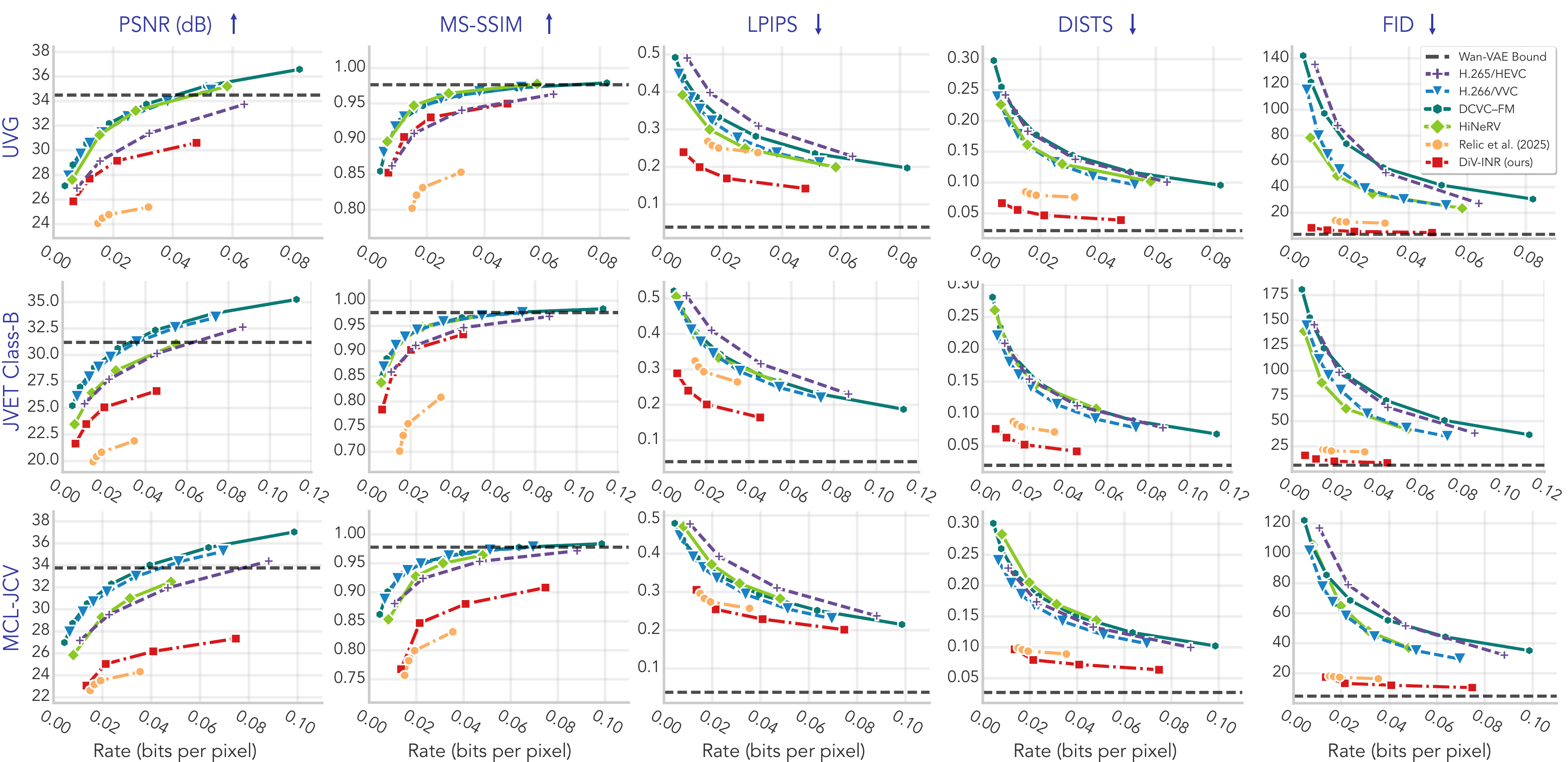}
  \caption{\textbf{Rate-distortion curves on UVG, JVET-B, and MCL-JCV.}
    Our approach (DiV-INR) is compared with traditional codecs (H.265/HEVC~\cite{hevc}, H.266/VVC~\cite{vvcoverview}), INR or neural video
    codecs (HiNeRV~\cite{kwanHiNeRVVideoCompression2024}, DCVC-FM~\cite{li2024neural}), and diffusion-based generative codecs (Relic~\etal~\cite{relic2025spatiotemporal}) using PSNR, MS-SSIM, LPIPS,
    DISTS, and FID across matched bitrates, highlighting the perceptual gains delivered by our
    INR-conditioned video diffusion approach.}
  \label{fig:rate-distortion-curves}
\end{figure*}

\subsection{Experimental setup}
\paragraph{Implementation details.}
We employ the distilled Wan2.1~\cite{wan2025wanopenadvancedlargescale} (1.3B parameters, $30$ DiT
blocks) as our video diffusion backbone, operating on 16-channel latents from a 3D causal VAE with
$4\times8\times8$ spatio-temporal downsampling. The DiT is conditioned on the INR without the
cross-attention mechanism (CLIP or textual prompt) to simplify the compression pipeline. For
INR, we use the HiNeRV~\cite{kwanHiNeRVVideoCompression2024} architecture, compressed via
15\% magnitude-based pruning and 6-bit quantization. NOLA adapters ($r{=}64$, $b{=}500$ bases) are
injected into attention output projections and feed-forward layers.

We train on $25$ frame GoPs on a single NVIDIA RTX 4090~(24GB) using mixed-precision
(\texttt{bfloat16}) and gradient checkpointing.
The three-stage schedule (dense, pruning-aware, quantization-aware) spans $480$ epochs per sequence. As computational cost is primarily dominated by the diffusion model, training time is not significantly affected by the INR capacity, yielding stable training duration across all bitrate points.
Adapter size primarily affects training: although raising the number of bases matrices improves perceptual quality, it also increases the computational cost of training more significantly compared to the INR capacity.
Due to the diminishing returns from increasing the number of bases matrices (see ablation studies in Sec.~\ref{subsec:ablation-inr-peft}) we cap the basis budget at $b{=}500$ to keep training practical while leaving inference cost flat via adapter merging.

\paragraph{Datasets and baselines.} \label{sec:datasets-and-baselines}
We evaluate our INR-based adaptive conditioning approach on standard video compression benchmarks:
UVG~\cite{uvg_dataset} (7 sequences), JVET Class-B~\cite{hevc} (5 sequences), and
MCL-JCV~\cite{wang2016mcl} (30 sequences).
We compare against H.265/HEVC~\cite{hevc}, H.266/VVC~\cite{vvcoverview},
DCVC-FM~\cite{li2024neural}, Relic~\etal~\cite{relic2025spatiotemporal}, and
HiNeRV~\cite{kwanHiNeRVVideoCompression2024}.
For HEVC, we use FFmpeg~\cite{ffmpeg2025} x265 v3.5 with the \texttt{veryslow} preset in 4:4:4 and GoP of 16.
For VVC, we use VTM-23.11~\cite{jvet_vtm} reference software in low-delay~P configuration with an intra period of 16.
All sequences are downsampled to $1024{\times}576$ to match the diffusion model's pre-training
distribution and 24\,GB GPU constraints, ensuring a fair comparison across all methods at the same
resolution. We report distortion metrics (PSNR and MS-SSIM) and perceptual metrics (LPIPS~\cite{zhang2018perceptual},
DISTS~\cite{ding2020iqa}, FID~\cite{heusel2018gans}) across multiple bitrate points.

\subsection{Results and Discussion}

\paragraph{Quantitative evaluation.}
\cref{tab:bd-rate-gains} and \cref{fig:rate-distortion-curves} present comprehensive results for DiV-INR
in terms of BD-metric summaries and rate-distortion curves, respectively. The rate-distortion plots in
\cref{fig:rate-distortion-curves} reveal that our method yields LPIPS, DISTS, and FID traces that
stay strictly below all baselines across $0.005{-}0.05$~bpp on UVG, JVET-B, and MCL-JCV while
the BD-metric summary in \cref{tab:bd-rate-gains} quantifies these gaps. Relative to the
strongest conventional codec, H.266/VVC~\cite{vvcoverview}, we improve BD-LPIPS by
$0.131/0.160/0.052$ (UVG/JVET-B/MCL-JCV); relative to a strong learned codec,
DCVC-FM~\cite{li2024neural}, by $0.156/0.175/0.049$; and relative to the base INR architecture
HiNeRV~\cite{kwanHiNeRVVideoCompression2024}, by $0.108/0.165/0.082$ BD-LPIPS, alongside gains of
$0.106/0.120/0.094$ BD-DISTS and $40.84/71.90/35.58$ BD-FID, respectively.

Among our benchmark datasets, improvements are even more pronounced on JVET Class-B for sequences
with complex motion and high-frequency details. However, PSNR and MS-SSIM results display the expected
perception--distortion trade-off~\cite{blauRethinkingLossyCompression2019}: DiV-INR trails traditional codecs and learned codecs by at least
$1$~dB BD-PSNR and comparable margins in MS-SSIM, while remaining better than or competitive with
the next-best perceptual compression method, Relic~\etal~\cite{relic2025spatiotemporal}. This mirrors
the theoretical rate-distortion-perception frontier~\cite{blauRethinkingLossyCompression2019},
wherein enforcing stronger perceptual alignment induces slightly higher pixel-domain distortion.
As other methods primarily optimize for distortion-based objectives, our approach prioritizes
diffusion loss and realism, leading to the observed performance gap in PSNR and MS-SSIM while
achieving significantly better results in LPIPS, DISTS, and FID.

Overall, the consistent LPIPS/DISTS/FID superiority in \cref{fig:rate-distortion-curves} and their
aggregate BD-metric advantages in \cref{tab:bd-rate-gains} validate that INR-conditioned diffusion
successfully matches the statistics of natural videos at bit budgets where conventional codecs
collapse in perception.

\begin{figure*}[t]
  \centering
  \includegraphics[width=\linewidth]{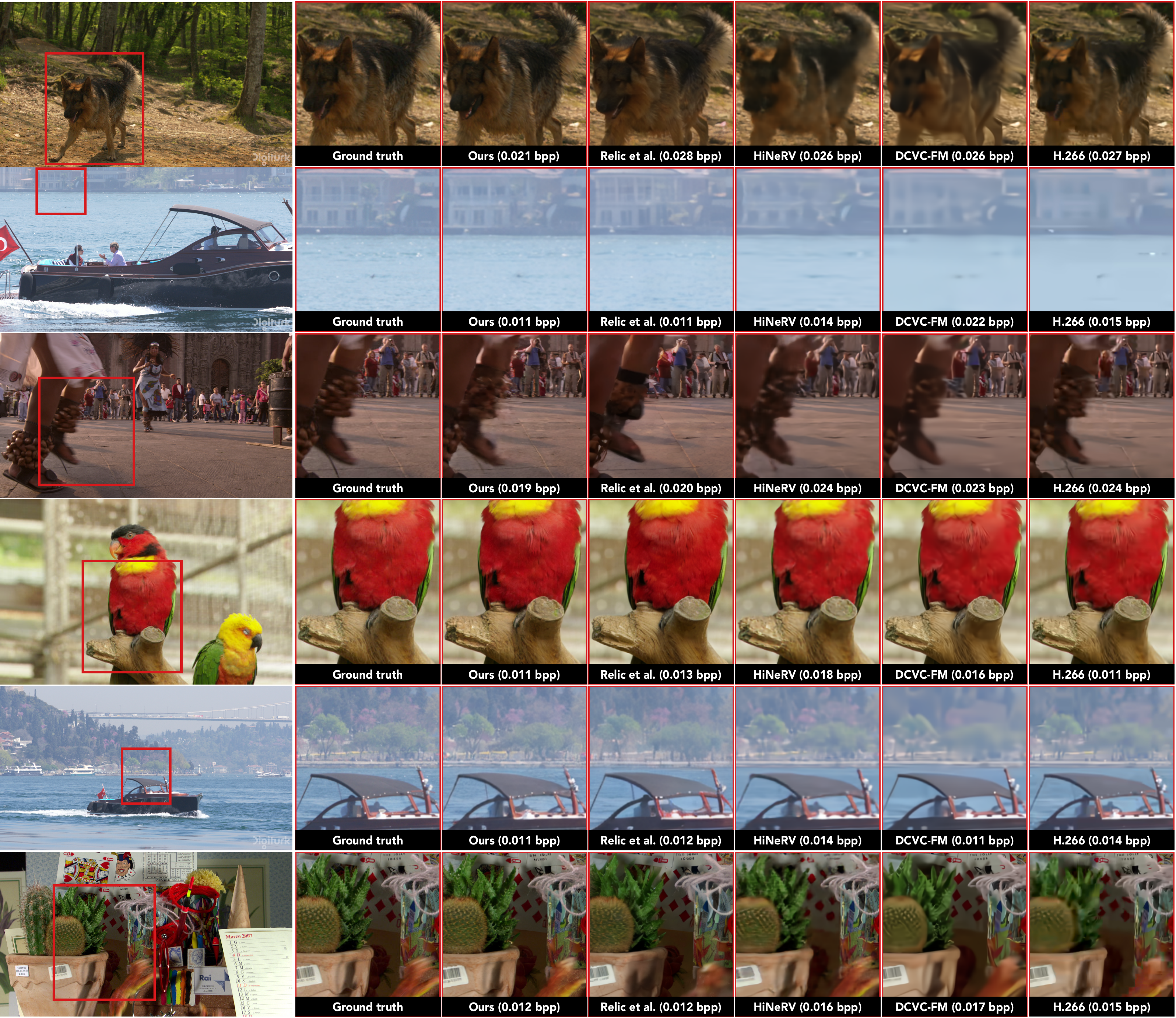}
  \caption{\textbf{Qualitative comparison on UVG, JVET-B, and MCL-JCV.}
    Representative crops show that our approach preserves high-frequency structure that
    Relic~\etal~\cite{relic2025spatiotemporal}, HiNeRV~\cite{kwanHiNeRVVideoCompression2024},
    DCVC-FM~\cite{li2024neural}, and VVC~\cite{vvcoverview} lose at similar bitrates.
    Best viewed digitally.}
  \label{fig:qualitative-comparison}
\end{figure*}

\paragraph{Qualitative evaluation.}
Qualitative comparisons in \cref{fig:qualitative-comparison} reveal the distinct characteristics of
our approach.
On \texttt{ShakeNDry} (high-frequency motion) and \texttt{YachtRide} (rapid camera movement), our method maintains sharp texture details where traditional and per-pixel distortion-oriented
codecs produce motion blur and blocking artifacts.
Similarly, \texttt{RitualDance} demonstrates improved frame-to-frame consistency in fine
details such as dancer movements and background textures. Our method generates plausible
high-frequency details in complex scenes (building textures, crowd dynamics) that are typically
lost in conventional codecs at extreme compression ratios.

\paragraph{Parameter overhead and bitrate breakdown.}
Each representation (INR + PEFT) contains ${\sim}100\text{K}{-}2.5\text{M}$ quantized INR
parameters for conditioning plus $91$K PEFT parameters, which compress to $0.005{-}0.05$ bpp across
our benchmarks in \cref{sec:datasets-and-baselines}.
Accordingly, $77{-}97\%$ of the bitrate payload is devoted to the INR, while PEFT adapters contribute the remaining $23-3\%$.
Because the INR alone specifies the video, we
avoid external keyframes and their GoP overlap overhead.

\paragraph{Training and inference complexity.}
DiV-INR targets \emph{offline, archi\-val} extreme compression where encoding cost is amortized and perceptual fidelity is the priority. 
The three-stage curriculum completes in ${\sim}15$ hours per video, requiring $21$\,GB peak VRAM.
Inference uses $20$ denoising steps with UniPC sampling~\cite{zhao2023unipcunifiedpredictorcorrectorframework} at ${\sim}1$ FPS (${\sim}12$\,GB peak VRAM).
Compared to Relic~\etal~\cite{relic2025spatiotemporal},
our decoding is significantly faster ($<0.1$ FPS for Relic~\etal) and fits within consumer VRAM during both training and inference.
While current diffusion-based decoding is not real-time, upcoming literature in causal video generation~\cite{yinSlowBidirectionalFast2025,huangSelfForcingBridging2025}, model distillation and efficient attention~(\eg, TurboDiffusion~\cite{turbodiffusion}, SageAttention~\cite{zhang2026sageattention3microscalingfp4attention}, VSA~\cite{zhang2025vsafastervideodiffusion}) can be directly integrated to reach practical frame rates.

\begin{figure}[hbtp]
  \centering
  \includegraphics[width=\linewidth]{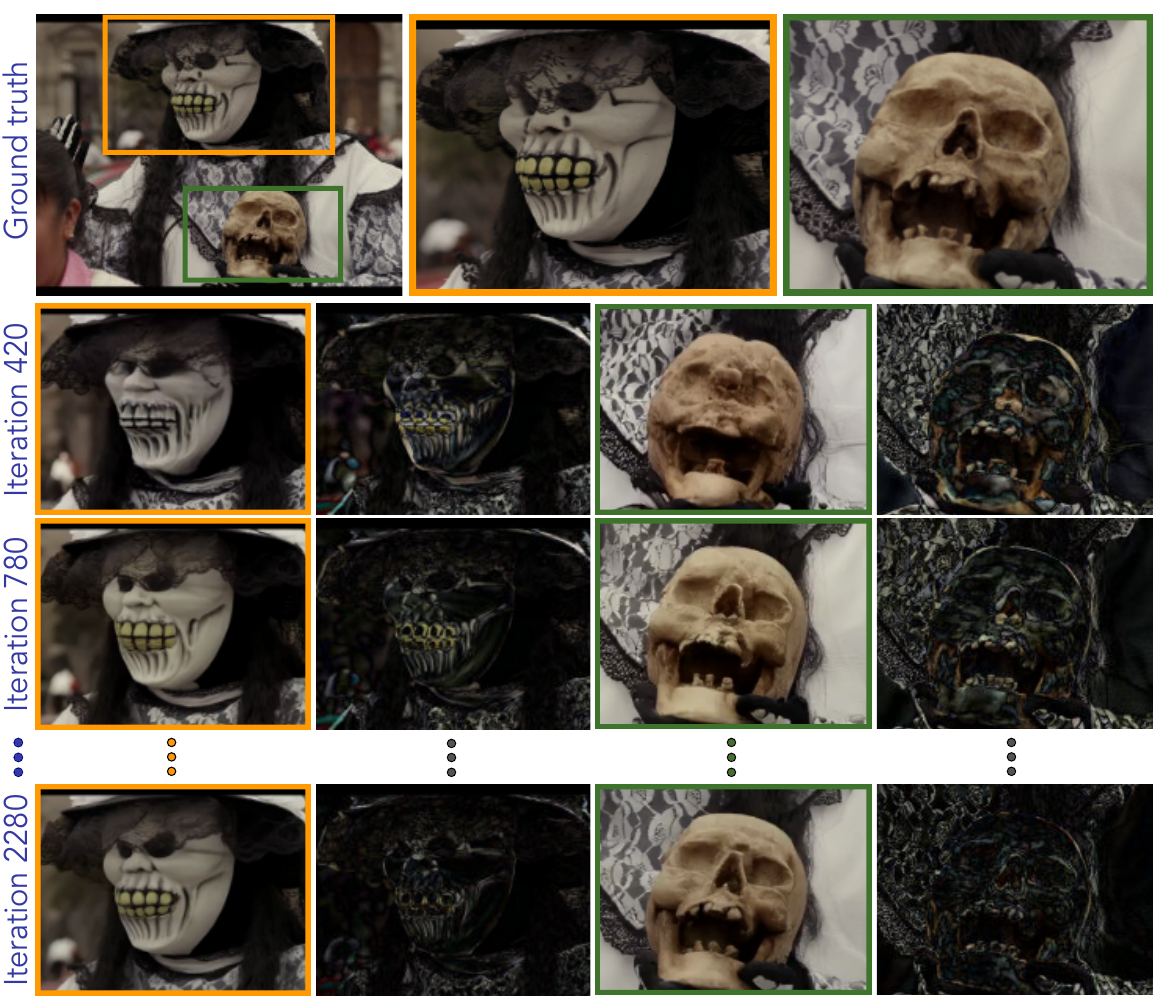}
  \caption{\textbf{Training dynamics for DiV-INR.}
    Snapshots across training iterations illustrate the semantic-first convergence at $0.019$ bpp:
    early samples match scene layout before textures, while later iterations align closely with
    ground truth. \label{fig:training_evolution}}
\end{figure}

\paragraph{Hierarchical convergence.} \label{sec:hierarchical-convergence}
\cref{fig:training_evolution} visualizes an example of training progression for our joint INR-adapter
optimization.
We observe a common hierarchical convergence pattern: the model initially generates semantically
coherent but visually distinct content~(correct object categories, scene layout, rough spatial
arrangement) before achieving pixel-level fidelity.
Early training stages~(e.g., iteration 420) produce outputs with significant appearance variations,
such as people with different features and paintings that are plausible yet different, while
maintaining semantic consistency with the ground-truth.

This semantic-to-visual refinement reveals fundamental insights into how generative models leverage
conditioning information.
The diffusion prior enables robust perceptual-oriented compression even with imperfect conditioning
signals during early training, potentially allowing more aggressive compression ratios than purely
reconstructive methods.
This progression suggests that early stopping strategies based on semantic consistency metrics
could reduce computational requirements while maintaining perceptual quality.

\subsection{Ablation Studies}
\label{subsec:ablation-inr-peft}

\begin{figure}[htbp]
  \centering
  \begin{subfigure}{0.5\linewidth}
    \centering
    \includegraphics[width=\linewidth]{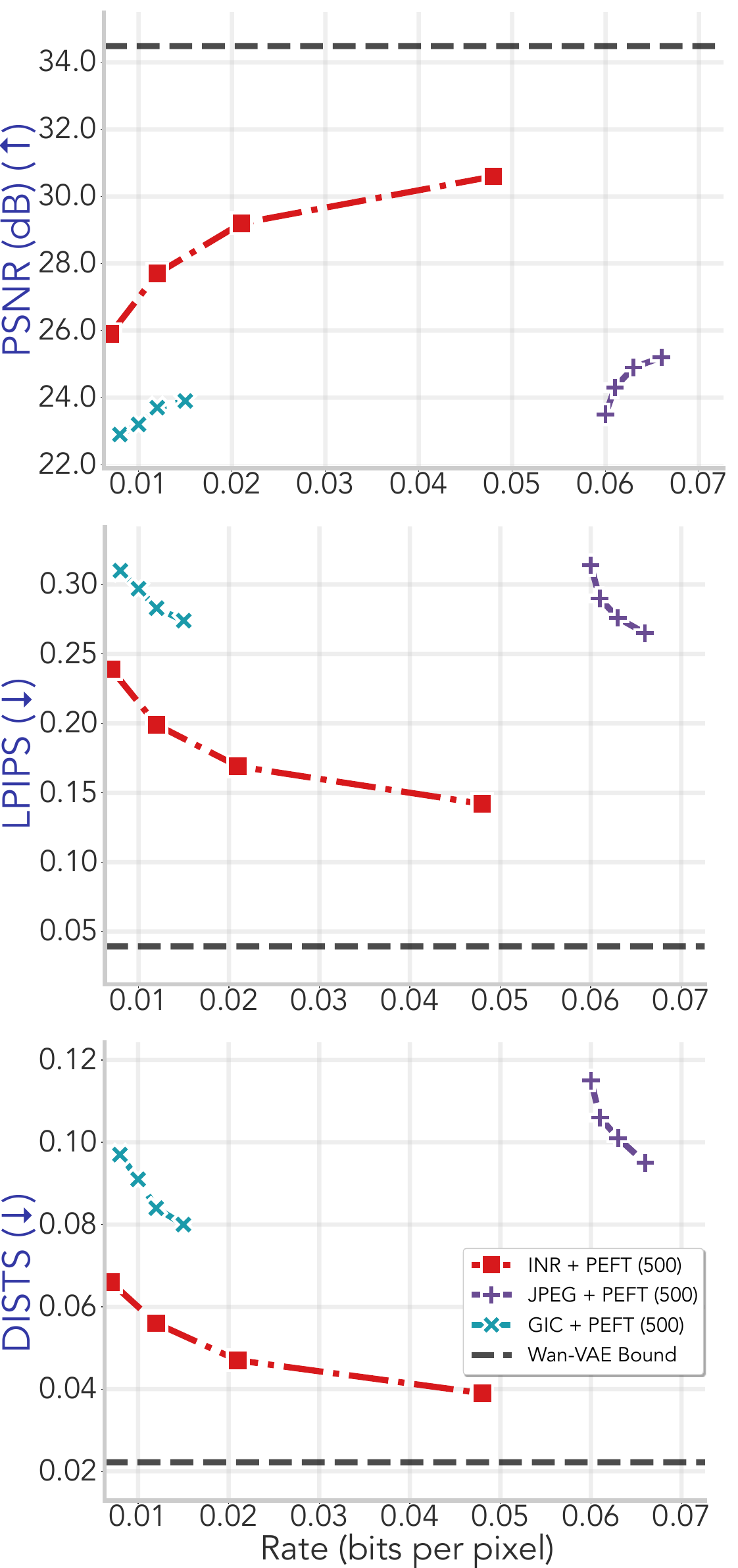}
    \captionsetup{width=0.9\linewidth}
    \caption{Impact of using INR vs. keyframes for conditioning.}
    \label{fig:ablation-inr-peft-a}
  \end{subfigure}%
  \begin{subfigure}{0.5\linewidth}
    \centering
    \includegraphics[width=\linewidth]{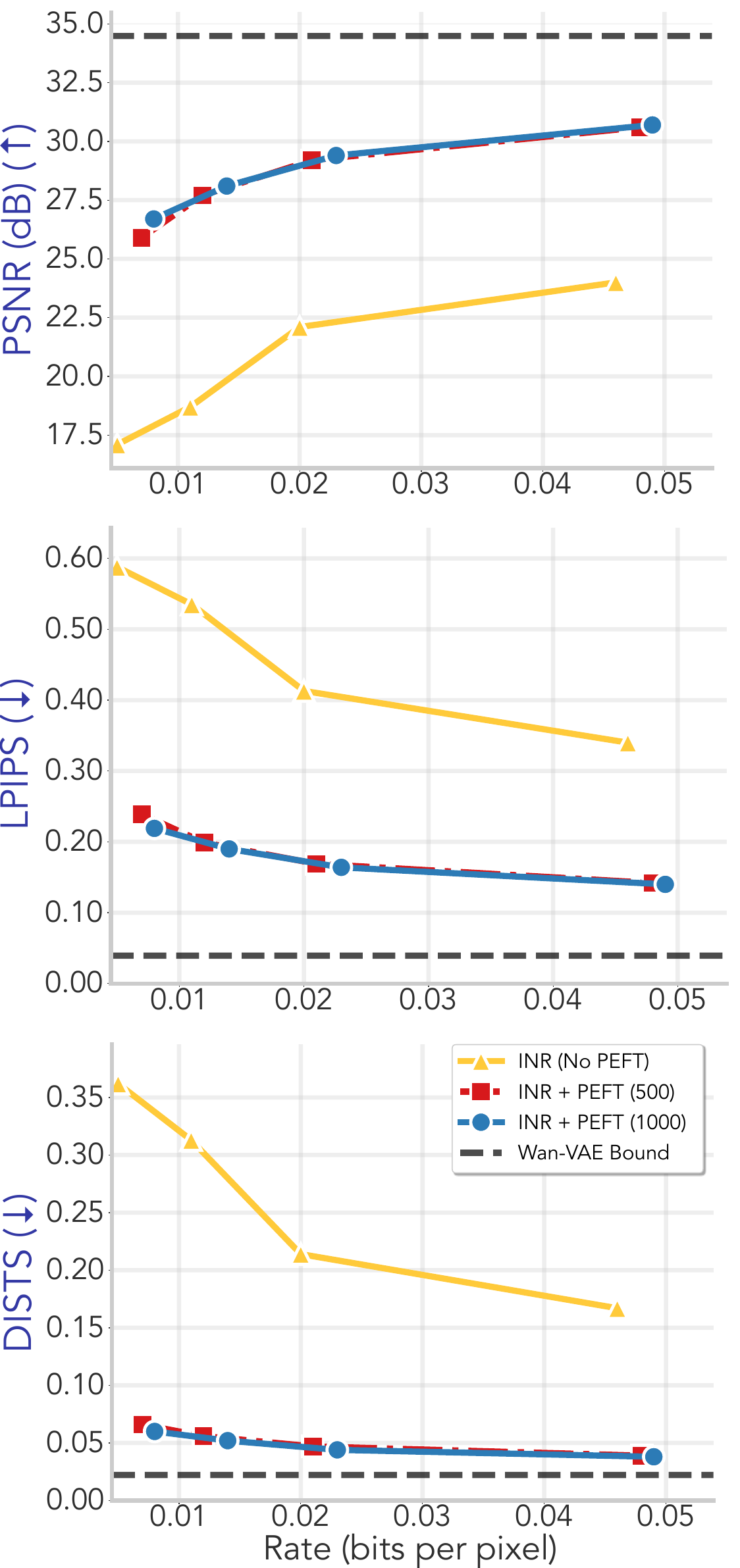}
    \captionsetup{width=0.9\linewidth}
    \caption{Impact of PEFT and its number of basis matrices.}
    \label{fig:ablation-inr-peft-b}
  \end{subfigure}
  \caption{\textbf{Ablation of INR conditioning and PEFT adaptation on UVG.}
    Enabling the INR (a) and progressively richer NOLA adapters boosts perceptual quality (b) while
    tracking the associated bitrate budget. However, increasing number of NOLA bases from $500$
    to $1000$ returns only marginal improvements in perceptual quality.}
  \label{fig:ablation-inr-peft}
\end{figure}

\paragraph{INR conditioning vs. intra-coded keyframes.}
\cref{fig:ablation-inr-peft-a} shows how rate-distortion performance varies when using INR conditioning versus intra-coded keyframes.
JPEG compressed keyframes incur a high bit cost, and the resulting method cannot achieve as low bitrates or maintain as high quality as our proposal.
Using a generative image codec~\cite{relic2025bridging}~(referred as \emph{GIC} in \cref{fig:ablation-inr-peft-a}) helps bridge the gap to low rate video compression.
However, it results in worse visual quality across all metrics compared to our proposal.

\paragraph{Impact of PEFT size.}
We ablate the effect of the PEFT adapter by performing experiments with a varying number of basis vectors, shown in \cref{fig:ablation-inr-peft-b}.
Omitting PEFT adapters entirely reduces rate-LPIPS performance by up to $0.239$, indicating that instance-specific finetuning of the diffusion model is necessary to fully exploit the conditioning information encoded in the INR.
However, we observe diminishing returns as the number of bases increases, with a LPIPS improvement of only $0.02$ between the 500 and 1000 basis variants at comparable bitrates.
As higher bases require longer training times, we choose 500 basis vectors to achieve a balance between compression performance and compute efficiency.

\paragraph{Impact of adaptive conditioning masks.} We evaluate the contribution of learned
spatiotemporal masks $\mM$ by comparing against a baseline where masks are replaced with uniform
confidence ($M{=}1$). This degradation in conditioning degrades BD-PSNR by $0.056$ dB and
BD-LPIPS by $0.003$ on UVG, confirming that learned uncertainty weighting is crucial for
effective conditioning especially in regions with complex motion.

\paragraph{Impact of alternative video diffusion models.} Our approach is structurally
backbone-agnostic, indicating that improvements in generative model translates to better compression.
In addition, while one might suspect that gains come only from the Wan2.1 backbone, swapping Wan2.1~\cite{wan2025wanopenadvancedlargescale}
into the Relic~\etal~\cite{relic2025spatiotemporal} pipeline actually degrades performance~(e.g., BD-PSNR 
drops by an additional $0.89$ dB compared to their SVD-based result).
This result confirms that our INR+PEFT integration is the primary quality driver.

\section{Conclusion}
\label{sec:conclusion}

We presented a novel framework for extreme low-bitrate video compression that combines implicit
neural representations with pre-trained video diffusion models through learned adaptive
conditioning.
Our approach demonstrates that compact neural representations can simultaneously serve as efficient
video encodings and optimized conditioning signals for generative models, enabling
instance-specific adaptation while preserving rich priors from foundation models.

Through systematic evaluation on established benchmarks, we validated substantial improvements in
perceptual metrics and qualitative performance at compression ratios where conventional codecs
struggle.
Results in \cref{sec:results} show consistent LPIPS/DISTS/FID gains over HEVC~\cite{hevc},
VVC~\cite{vvcoverview}, and recent neural codecs (Relic~\etal~\cite{relic2025spatiotemporal},
HiNeRV~\cite{kwanHiNeRVVideoCompression2024}, DCVC-FM~\cite{li2024neural}) on UVG, JVET-B, and
MCL-JCV, confirming that INR latent-and-mask predictions and PEFT adapters can be treated as representations decoded with video diffusion models, which serve as inherent video decoders.
Finally, our analysis reveals a semantic-first convergence pattern in~\cref{sec:hierarchical-convergence} which explains the robustness of perceptual quality at ultra-low bitrates and motivates future semantic-aware stopping criteria and conditioning schedules.

\bibliographystyle{ACM-Reference-Format}
\bibliography{main}


\begin{thebibliography}{43}


\ifx \showCODEN    \undefined \def \showCODEN     #1{\unskip}     \fi
\ifx \showISBNx    \undefined \def \showISBNx     #1{\unskip}     \fi
\ifx \showISBNxiii \undefined \def \showISBNxiii  #1{\unskip}     \fi
\ifx \showISSN     \undefined \def \showISSN      #1{\unskip}     \fi
\ifx \showLCCN     \undefined \def \showLCCN      #1{\unskip}     \fi
\ifx \shownote     \undefined \def \shownote      #1{#1}          \fi
\ifx \showarticletitle \undefined \def \showarticletitle #1{#1}   \fi
\ifx \showURL      \undefined \def \showURL       {\relax}        \fi
\providecommand\bibfield[2]{#2}
\providecommand\bibinfo[2]{#2}
\providecommand\natexlab[1]{#1}
\providecommand\showeprint[2][]{arXiv:#2}

\bibitem[Blattmann et~al\mbox{.}(2023)]%
        {blattmann2023stablevideodiffusionscaling}
\bibfield{author}{\bibinfo{person}{Andreas Blattmann}, \bibinfo{person}{Tim
  Dockhorn}, \bibinfo{person}{Sumith Kulal}, \bibinfo{person}{Daniel
  Mendelevitch}, \bibinfo{person}{Maciej Kilian}, \bibinfo{person}{Dominik
  Lorenz}, \bibinfo{person}{Yam Levi}, \bibinfo{person}{Zion English},
  \bibinfo{person}{Vikram Voleti}, \bibinfo{person}{Adam Letts},
  \bibinfo{person}{Varun Jampani}, {and} \bibinfo{person}{Robin Rombach}.}
  \bibinfo{year}{2023}\natexlab{}.
\newblock \bibinfo{title}{Stable Video Diffusion: Scaling Latent Video
  Diffusion Models to Large Datasets}.
\newblock
\showeprint[arxiv]{2311.15127}~[cs. CV]
\urldef\tempurl%
\url{https://arxiv.org/abs/2311.15127}
\showURL{%
\tempurl}


\bibitem[Blau and Michaeli(2019)]%
        {blauRethinkingLossyCompression2019}
\bibfield{author}{\bibinfo{person}{Yochai Blau} {and} \bibinfo{person}{Tomer
  Michaeli}.} \bibinfo{year}{2019}\natexlab{}.
\newblock \bibinfo{title}{Rethinking {{Lossy Compression}}: {{The
  Rate-Distortion-Perception Tradeoff}}}.
\newblock
\showeprint[arxiv]{1901.07821}~[cs]
\href{https://doi.org/10.48550/arXiv.1901.07821}{doi:\nolinkurl{10.48550/arXiv.1901.07821}}


\bibitem[Bross et~al\mbox{.}(2021)]%
        {vvcoverview}
\bibfield{author}{\bibinfo{person}{Benjamin Bross}, \bibinfo{person}{Ye-Kui
  Wang}, \bibinfo{person}{Yan Ye}, \bibinfo{person}{Shan Liu},
  \bibinfo{person}{Jianle Chen}, \bibinfo{person}{Gary~J. Sullivan}, {and}
  \bibinfo{person}{Jens-Rainer Ohm}.} \bibinfo{year}{2021}\natexlab{}.
\newblock \showarticletitle{Overview of the Versatile Video Coding (VVC)
  Standard and its Applications}.
\newblock \bibinfo{journal}{\emph{IEEE Transactions on Circuits and Systems for
  Video Technology}} \bibinfo{volume}{31}, \bibinfo{number}{10}
  (\bibinfo{year}{2021}), \bibinfo{pages}{3736--3764}.
\newblock
\href{https://doi.org/10.1109/TCSVT.2021.3101953}{doi:\nolinkurl{10.1109/TCSVT.2021.3101953}}


\bibitem[Chen et~al\mbox{.}(2023a)]%
        {chen2023hnervhybridneuralrepresentation}
\bibfield{author}{\bibinfo{person}{Hao Chen}, \bibinfo{person}{Matt Gwilliam},
  \bibinfo{person}{Ser-Nam Lim}, {and} \bibinfo{person}{Abhinav Shrivastava}.}
  \bibinfo{year}{2023}\natexlab{a}.
\newblock \bibinfo{title}{HNeRV: A Hybrid Neural Representation for Videos}.
\newblock
\showeprint[arxiv]{2304.02633}~[cs.CV]
\urldef\tempurl%
\url{https://arxiv.org/abs/2304.02633}
\showURL{%
\tempurl}


\bibitem[Chen et~al\mbox{.}(2021)]%
        {chenNeRVNeuralRepresentations2021}
\bibfield{author}{\bibinfo{person}{Hao Chen}, \bibinfo{person}{Bo He},
  \bibinfo{person}{Hanyu Wang}, \bibinfo{person}{Yixuan Ren},
  \bibinfo{person}{Ser-Nam Lim}, {and} \bibinfo{person}{Abhinav Shrivastava}.}
  \bibinfo{year}{2021}\natexlab{}.
\newblock \bibinfo{title}{{{NeRV}}: {{Neural Representations}} for {{Videos}}}.
\newblock
\showeprint[arxiv]{2110.13903}~[cs]
\href{https://doi.org/10.48550/arXiv.2110.13903}{doi:\nolinkurl{10.48550/arXiv.2110.13903}}


\bibitem[Chen et~al\mbox{.}(2023b)]%
        {zhenghao2023neural}
\bibfield{author}{\bibinfo{person}{Zhenghao Chen}, \bibinfo{person}{Lucas
  Relic}, \bibinfo{person}{Roberto Azevedo}, \bibinfo{person}{Yang Zhang},
  \bibinfo{person}{Markus Gross}, \bibinfo{person}{Dong Xu},
  \bibinfo{person}{Luping Zhou}, {and} \bibinfo{person}{Christopher Schroers}.}
  \bibinfo{year}{2023}\natexlab{b}.
\newblock \showarticletitle{Neural Video Compression with Spatio-Temporal
  Cross-Covariance Transformers}. In \bibinfo{booktitle}{\emph{Proceedings of
  the 31st ACM International Conference on Multimedia}} (Ottawa ON, Canada)
  \emph{(\bibinfo{series}{MM '23})}. \bibinfo{publisher}{Association for
  Computing Machinery}, \bibinfo{address}{New York, NY, USA},
  \bibinfo{pages}{8543–8551}.
\newblock
\showISBNx{9798400701085}
\href{https://doi.org/10.1145/3581783.3611960}{doi:\nolinkurl{10.1145/3581783.3611960}}


\bibitem[Ding et~al\mbox{.}(2020)]%
        {ding2020iqa}
\bibfield{author}{\bibinfo{person}{Keyan Ding}, \bibinfo{person}{Kede Ma},
  \bibinfo{person}{Shiqi Wang}, {and} \bibinfo{person}{Eero~P. Simoncelli}.}
  \bibinfo{year}{2020}\natexlab{}.
\newblock \showarticletitle{Image Quality Assessment: Unifying Structure and
  Texture Similarity}.
\newblock \bibinfo{journal}{\emph{CoRR}}  \bibinfo{volume}{abs/2004.07728}
  (\bibinfo{year}{2020}).
\newblock
\urldef\tempurl%
\url{https://arxiv.org/abs/2004.07728}
\showURL{%
\tempurl}


\bibitem[Dupont et~al\mbox{.}(2021)]%
        {dupont2021coin}
\bibfield{author}{\bibinfo{person}{Emilien Dupont}, \bibinfo{person}{Adam
  Golinski}, \bibinfo{person}{Milad Alizadeh}, \bibinfo{person}{Yee~Whye Teh},
  {and} \bibinfo{person}{Arnaud Doucet}.} \bibinfo{year}{2021}\natexlab{}.
\newblock \showarticletitle{{COIN}: {CO}mpression with Implicit Neural
  representations}. In \bibinfo{booktitle}{\emph{Neural Compression: From
  Information Theory to Applications -- Workshop @ ICLR 2021}}.
\newblock
\urldef\tempurl%
\url{https://openreview.net/forum?id=yekxhcsVi4}
\showURL{%
\tempurl}


\bibitem[Fan et~al\mbox{.}(2021)]%
        {fanTrainingQuantizationNoise2021}
\bibfield{author}{\bibinfo{person}{Angela Fan}, \bibinfo{person}{Pierre Stock},
  \bibinfo{person}{Benjamin Graham}, \bibinfo{person}{Edouard Grave},
  \bibinfo{person}{Remi Gribonval}, \bibinfo{person}{Herve Jegou}, {and}
  \bibinfo{person}{Armand Joulin}.} \bibinfo{year}{2021}\natexlab{}.
\newblock \bibinfo{title}{Training with Quantization Noise for Extreme Model
  Compression}.
\newblock
\showeprint[arxiv]{2004.07320}~[cs]
\href{https://doi.org/10.48550/arXiv.2004.07320}{doi:\nolinkurl{10.48550/arXiv.2004.07320}}


\bibitem[{FFmpeg Developers}(2025)]%
        {ffmpeg2025}
\bibfield{author}{\bibinfo{person}{{FFmpeg Developers}}.}
  \bibinfo{year}{2025}\natexlab{}.
\newblock \bibinfo{title}{FFmpeg documentation -- a complete, cross-platform
  solution to record, convert and stream audio and video}.
\newblock \bibinfo{howpublished}{\url{https://ffmpeg.org/documentation.html}}.
\newblock
\newblock
\shownote{Version 7.1 (git commit <abcd123>), accessed 26 Jun 2025}.


\bibitem[Gao et~al\mbox{.}(2025)]%
        {gao2025givicgenerativeimplicitvideo}
\bibfield{author}{\bibinfo{person}{Ge Gao}, \bibinfo{person}{Siyue Teng},
  \bibinfo{person}{Tianhao Peng}, \bibinfo{person}{Fan Zhang}, {and}
  \bibinfo{person}{David Bull}.} \bibinfo{year}{2025}\natexlab{}.
\newblock \bibinfo{title}{GIViC: Generative Implicit Video Compression}.
\newblock
\showeprint[arxiv]{2503.19604}~[eess.IV]
\urldef\tempurl%
\url{https://arxiv.org/abs/2503.19604}
\showURL{%
\tempurl}


\bibitem[Gomes et~al\mbox{.}(2023)]%
        {gomesVideoCompressionEntropyConstrained2023a}
\bibfield{author}{\bibinfo{person}{Carlos Gomes}, \bibinfo{person}{Roberto
  Azevedo}, {and} \bibinfo{person}{Christopher Schroers}.}
  \bibinfo{year}{2023}\natexlab{}.
\newblock \showarticletitle{Video {{Compression}} with {{Entropy-Constrained
  Neural Representations}}}. In \bibinfo{booktitle}{\emph{2023 {{IEEE}}/{{CVF
  Conference}} on {{Computer Vision}} and {{Pattern Recognition}} ({{CVPR}})}}.
  \bibinfo{publisher}{IEEE}, \bibinfo{address}{Vancouver, BC, Canada},
  \bibinfo{pages}{18497--18506}.
\newblock
\showISBNx{979-8-3503-0129-8}
\href{https://doi.org/10.1109/CVPR52729.2023.01774}{doi:\nolinkurl{10.1109/CVPR52729.2023.01774}}


\bibitem[Han et~al\mbox{.}(2021)]%
        {han2021technicaloverviewav1}
\bibfield{author}{\bibinfo{person}{Jingning Han}, \bibinfo{person}{Bohan Li},
  \bibinfo{person}{Debargha Mukherjee}, \bibinfo{person}{Ching-Han Chiang},
  \bibinfo{person}{Adrian Grange}, \bibinfo{person}{Cheng Chen},
  \bibinfo{person}{Hui Su}, \bibinfo{person}{Sarah Parker},
  \bibinfo{person}{Sai Deng}, \bibinfo{person}{Urvang Joshi},
  \bibinfo{person}{Yue Chen}, \bibinfo{person}{Yunqing Wang},
  \bibinfo{person}{Paul Wilkins}, \bibinfo{person}{Yaowu Xu}, {and}
  \bibinfo{person}{James Bankoski}.} \bibinfo{year}{2021}\natexlab{}.
\newblock \bibinfo{title}{A Technical Overview of AV1}.
\newblock
\showeprint[arxiv]{2008.06091}~[eess.IV]
\urldef\tempurl%
\url{https://arxiv.org/abs/2008.06091}
\showURL{%
\tempurl}


\bibitem[Heusel et~al\mbox{.}(2018)]%
        {heusel2018gans}
\bibfield{author}{\bibinfo{person}{Martin Heusel}, \bibinfo{person}{Hubert
  Ramsauer}, \bibinfo{person}{Thomas Unterthiner}, \bibinfo{person}{Bernhard
  Nessler}, {and} \bibinfo{person}{Sepp Hochreiter}.}
  \bibinfo{year}{2018}\natexlab{}.
\newblock \bibinfo{title}{GANs Trained by a Two Time-Scale Update Rule Converge
  to a Local Nash Equilibrium}.
\newblock
\showeprint[arxiv]{1706.08500}~[cs.LG]
\urldef\tempurl%
\url{https://arxiv.org/abs/1706.08500}
\showURL{%
\tempurl}


\bibitem[Hu et~al\mbox{.}(2021)]%
        {huLoRALowRankAdaptation2021}
\bibfield{author}{\bibinfo{person}{Edward~J. Hu}, \bibinfo{person}{Yelong
  Shen}, \bibinfo{person}{Phillip Wallis}, \bibinfo{person}{Zeyuan
  {Allen-Zhu}}, \bibinfo{person}{Yuanzhi Li}, \bibinfo{person}{Shean Wang},
  \bibinfo{person}{Lu Wang}, {and} \bibinfo{person}{Weizhu Chen}.}
  \bibinfo{year}{2021}\natexlab{}.
\newblock \bibinfo{title}{{{LoRA}}: {{Low-Rank Adaptation}} of {{Large Language
  Models}}}.
\newblock
\showeprint[arxiv]{2106.09685}~[cs]
\href{https://doi.org/10.48550/arXiv.2106.09685}{doi:\nolinkurl{10.48550/arXiv.2106.09685}}


\bibitem[Huang et~al\mbox{.}(2025)]%
        {huangSelfForcingBridging2025}
\bibfield{author}{\bibinfo{person}{Xun Huang}, \bibinfo{person}{Zhengqi Li},
  \bibinfo{person}{Guande He}, \bibinfo{person}{Mingyuan Zhou}, {and}
  \bibinfo{person}{Eli Shechtman}.} \bibinfo{year}{2025}\natexlab{}.
\newblock \bibinfo{title}{Self {{Forcing}}: {{Bridging}} the {{Train-Test Gap}}
  in {{Autoregressive Video Diffusion}}}.
\newblock
\showeprint{2506.08009}~[cs]
\href{https://doi.org/10.48550/arXiv.2506.08009}{doi:\nolinkurl{10.48550/arXiv.2506.08009}}


\bibitem[{Joint Video Experts Team (JVET)}({[n.\,d.]})]%
        {jvet_vtm}
\bibfield{author}{\bibinfo{person}{{Joint Video Experts Team (JVET)}}.}
  \bibinfo{year}{[n.\,d.]}\natexlab{}.
\newblock \bibinfo{title}{VVC Test Model (VTM) Reference Software}.
\newblock \bibinfo{howpublished}{\url{https://jvet.hhi.fraunhofer.de/}}.
\newblock
\newblock
\shownote{Online; accessed 20 November 2025}.


\bibitem[Koohpayegani et~al\mbox{.}(2024)]%
        {koohpayeganiNOLACompressingLoRA2024}
\bibfield{author}{\bibinfo{person}{Soroush~Abbasi Koohpayegani},
  \bibinfo{person}{K.~L. Navaneet}, \bibinfo{person}{Parsa Nooralinejad},
  \bibinfo{person}{Soheil Kolouri}, {and} \bibinfo{person}{Hamed Pirsiavash}.}
  \bibinfo{year}{2024}\natexlab{}.
\newblock \bibinfo{title}{{{NOLA}}: {{Compressing LoRA}} Using {{Linear
  Combination}} of {{Random Basis}}}.
\newblock
\showeprint[arxiv]{2310.02556}~[cs]
\href{https://doi.org/10.48550/arXiv.2310.02556}{doi:\nolinkurl{10.48550/arXiv.2310.02556}}


\bibitem[Kwan et~al\mbox{.}(2024)]%
        {kwanHiNeRVVideoCompression2024}
\bibfield{author}{\bibinfo{person}{Ho~Man Kwan}, \bibinfo{person}{Ge Gao},
  \bibinfo{person}{Fan Zhang}, \bibinfo{person}{Andrew Gower}, {and}
  \bibinfo{person}{David Bull}.} \bibinfo{year}{2024}\natexlab{}.
\newblock \bibinfo{title}{{{HiNeRV}}: {{Video Compression}} with {{Hierarchical
  Encoding-based Neural Representation}}}.
\newblock
\showeprint[arxiv]{2306.09818}~[eess]
\href{https://doi.org/10.5555/3666122.3669299}{doi:\nolinkurl{10.5555/3666122.3669299}}


\bibitem[Li et~al\mbox{.}(2024b)]%
        {li2024extremevideocompressionpretrained}
\bibfield{author}{\bibinfo{person}{Bohan Li}, \bibinfo{person}{Yiming Liu},
  \bibinfo{person}{Xueyan Niu}, \bibinfo{person}{Bo Bai}, \bibinfo{person}{Lei
  Deng}, {and} \bibinfo{person}{Deniz Gündüz}.}
  \bibinfo{year}{2024}\natexlab{b}.
\newblock \bibinfo{title}{Extreme Video Compression with Pre-trained Diffusion
  Models}.
\newblock
\showeprint[arxiv]{2402.08934}~[eess.IV]
\urldef\tempurl%
\url{https://arxiv.org/abs/2402.08934}
\showURL{%
\tempurl}


\bibitem[Li et~al\mbox{.}(2021)]%
        {li2021Deep}
\bibfield{author}{\bibinfo{person}{Jiahao Li}, \bibinfo{person}{Bin Li}, {and}
  \bibinfo{person}{Yan Lu}.} \bibinfo{year}{2021}\natexlab{}.
\newblock \showarticletitle{Deep {{Contextual Video Compression}}}. In
  \bibinfo{booktitle}{\emph{Advances in {{Neural Information Processing
  Systems}}}}, Vol.~\bibinfo{volume}{34}. \bibinfo{publisher}{Curran
  Associates, Inc.}, \bibinfo{pages}{18114--18125}.
\newblock


\bibitem[Li et~al\mbox{.}(2024a)]%
        {li2024neural}
\bibfield{author}{\bibinfo{person}{Jiahao Li}, \bibinfo{person}{Bin Li}, {and}
  \bibinfo{person}{Yan Lu}.} \bibinfo{year}{2024}\natexlab{a}.
\newblock \showarticletitle{Neural Video Compression with Feature Modulation}.
  In \bibinfo{booktitle}{\emph{{IEEE/CVF} Conference on Computer Vision and
  Pattern Recognition, {CVPR} 2024, Seattle, WA, USA, June 17-21, 2024}}.
\newblock


\bibitem[Li et~al\mbox{.}(2022)]%
        {li2022enervexpediteneuralvideo}
\bibfield{author}{\bibinfo{person}{Zizhang Li}, \bibinfo{person}{Mengmeng
  Wang}, \bibinfo{person}{Huaijin Pi}, \bibinfo{person}{Kechun Xu},
  \bibinfo{person}{Jianbiao Mei}, {and} \bibinfo{person}{Yong Liu}.}
  \bibinfo{year}{2022}\natexlab{}.
\newblock \bibinfo{title}{E-NeRV: Expedite Neural Video Representation with
  Disentangled Spatial-Temporal Context}.
\newblock
\showeprint[arxiv]{2207.08132}~[cs.CV]
\urldef\tempurl%
\url{https://arxiv.org/abs/2207.08132}
\showURL{%
\tempurl}


\bibitem[Lipman et~al\mbox{.}(2023)]%
        {lipman2023fm}
\bibfield{author}{\bibinfo{person}{Yaron Lipman}, \bibinfo{person}{Ricky T.~Q.
  Chen}, \bibinfo{person}{Heli Ben-Hamu}, \bibinfo{person}{Maximilian Nickel},
  {and} \bibinfo{person}{Matt Le}.} \bibinfo{year}{2023}\natexlab{}.
\newblock \bibinfo{title}{Flow Matching for Generative Modeling}.
\newblock
\showeprint[arxiv]{2210.02747}~[cs.LG]
\urldef\tempurl%
\url{https://arxiv.org/abs/2210.02747}
\showURL{%
\tempurl}


\bibitem[Mercat et~al\mbox{.}(2020)]%
        {uvg_dataset}
\bibfield{author}{\bibinfo{person}{Alexandre Mercat}, \bibinfo{person}{Marko
  Viitanen}, {and} \bibinfo{person}{Jarno Vanne}.}
  \bibinfo{year}{2020}\natexlab{}.
\newblock \showarticletitle{UVG dataset: 50/120fps 4K sequences for video codec
  analysis and development}. In \bibinfo{booktitle}{\emph{Proceedings of the
  11th ACM Multimedia Systems Conference}} (Istanbul, Turkey)
  \emph{(\bibinfo{series}{MMSys '20})}. \bibinfo{publisher}{Association for
  Computing Machinery}, \bibinfo{address}{New York, NY, USA},
  \bibinfo{pages}{297–302}.
\newblock
\showISBNx{9781450368452}
\href{https://doi.org/10.1145/3339825.3394937}{doi:\nolinkurl{10.1145/3339825.3394937}}


\bibitem[Relic et~al\mbox{.}(2025a)]%
        {relic2024lossyimagecompression}
\bibfield{author}{\bibinfo{person}{Lucas Relic}, \bibinfo{person}{Roberto
  Azevedo}, \bibinfo{person}{Markus Gross}, {and} \bibinfo{person}{Christopher
  Schroers}.} \bibinfo{year}{2025}\natexlab{a}.
\newblock \showarticletitle{Lossy Image Compression with Foundation Diffusion
  Models}. In \bibinfo{booktitle}{\emph{Computer Vision -- {{ECCV}} 2024}},
  \bibfield{editor}{\bibinfo{person}{Ale{\v s} Leonardis},
  \bibinfo{person}{Elisa Ricci}, \bibinfo{person}{Stefan Roth},
  \bibinfo{person}{Olga Russakovsky}, \bibinfo{person}{Torsten Sattler}, {and}
  \bibinfo{person}{G{\"u}l Varol}} (Eds.). \bibinfo{publisher}{Springer Nature
  Switzerland}, \bibinfo{address}{Cham}, \bibinfo{pages}{303--319}.
\newblock
\showISBNx{978-3-031-73030-6}


\bibitem[Relic et~al\mbox{.}(2025b)]%
        {relic2025bridging}
\bibfield{author}{\bibinfo{person}{L. Relic}, \bibinfo{person}{R. Azevedo},
  \bibinfo{person}{Y. Zhang}, \bibinfo{person}{M. Gross}, {and}
  \bibinfo{person}{C. Schroers}.} \bibinfo{year}{2025}\natexlab{b}.
\newblock \showarticletitle{Bridging the Gap between Gaussian Diffusion Models
  and Universal Quantization for Image Compression}. In
  \bibinfo{booktitle}{\emph{CVPR}}.
\newblock


\bibitem[Relic et~al\mbox{.}(2025c)]%
        {relic2025spatiotemporal}
\bibfield{author}{\bibinfo{person}{Lucas Relic}, \bibinfo{person}{Andr\'e
  Emmenegger}, \bibinfo{person}{Roberto Azevedo}, \bibinfo{person}{Yang Zhang},
  \bibinfo{person}{Markus Gross}, {and} \bibinfo{person}{Christopher
  Schroers}.} \bibinfo{year}{2025}\natexlab{c}.
\newblock \showarticletitle{Spatiotemporal Diffusion Priors for Extreme Video
  Compression}. In \bibinfo{booktitle}{\emph{2025 Picture Coding
  Symposium~(PCS)}}. IEEE.
\newblock


\bibitem[Saethre et~al\mbox{.}(2024)]%
        {saethreCombiningFrameGOP2024}
\bibfield{author}{\bibinfo{person}{Jens~Eirik Saethre},
  \bibinfo{person}{Roberto Azevedo}, {and} \bibinfo{person}{Christopher
  Schroers}.} \bibinfo{year}{2024}\natexlab{}.
\newblock \showarticletitle{Combining {{Frame}} and {{GOP Embeddings}} for
  {{Neural Video Representation}}}. In \bibinfo{booktitle}{\emph{2024
  {{IEEE}}/{{CVF Conference}} on {{Computer Vision}} and {{Pattern
  Recognition}} ({{CVPR}})}}. \bibinfo{publisher}{IEEE},
  \bibinfo{address}{Seattle, WA, USA}, \bibinfo{pages}{9253--9263}.
\newblock
\showISBNx{979-8-3503-5300-6}
\href{https://doi.org/10.1109/CVPR52733.2024.00884}{doi:\nolinkurl{10.1109/CVPR52733.2024.00884}}


\bibitem[Sitzmann et~al\mbox{.}(2020)]%
        {sitzmannImplicitNeuralRepresentations2020}
\bibfield{author}{\bibinfo{person}{Vincent Sitzmann}, \bibinfo{person}{Julien
  N.~P. Martel}, \bibinfo{person}{Alexander~W. Bergman},
  \bibinfo{person}{David~B. Lindell}, {and} \bibinfo{person}{Gordon
  Wetzstein}.} \bibinfo{year}{2020}\natexlab{}.
\newblock \bibinfo{title}{Implicit {{Neural Representations}} with {{Periodic
  Activation Functions}}}.
\newblock
\showeprint[arxiv]{2006.09661}~[cs]
\href{https://doi.org/10.48550/arXiv.2006.09661}{doi:\nolinkurl{10.48550/arXiv.2006.09661}}


\bibitem[Sze et~al\mbox{.}(2014)]%
        {hevc}
\bibfield{author}{\bibinfo{person}{Vivienne Sze}, \bibinfo{person}{Madhukar
  Budagavi}, {and} \bibinfo{person}{Gary~J. Sullivan}.}
  \bibinfo{year}{2014}\natexlab{}.
\newblock \bibinfo{booktitle}{\emph{High Efficiency Video Coding (HEVC):
  Algorithms and Architectures}}.
\newblock \bibinfo{publisher}{Springer}.
\newblock
\showISBNx{978-3-319-06894-7}
\href{https://doi.org/10.1007/978-3-319-06895-4}{doi:\nolinkurl{10.1007/978-3-319-06895-4}}


\bibitem[Wan et~al\mbox{.}(2025)]%
        {wan2025wanopenadvancedlargescale}
\bibfield{author}{\bibinfo{person}{Team Wan}, \bibinfo{person}{Ang Wang},
  \bibinfo{person}{Baole Ai}, \bibinfo{person}{Bin Wen},
  \bibinfo{person}{Chaojie Mao}, \bibinfo{person}{Chen-Wei Xie},
  \bibinfo{person}{Di Chen}, \bibinfo{person}{Feiwu Yu},
  \bibinfo{person}{Haiming Zhao}, \bibinfo{person}{Jianxiao Yang},
  \bibinfo{person}{Jianyuan Zeng}, \bibinfo{person}{Jiayu Wang},
  \bibinfo{person}{Jingfeng Zhang}, \bibinfo{person}{Jingren Zhou},
  \bibinfo{person}{Jinkai Wang}, \bibinfo{person}{Jixuan Chen},
  \bibinfo{person}{Kai Zhu}, \bibinfo{person}{Kang Zhao}, \bibinfo{person}{Keyu
  Yan}, \bibinfo{person}{Lianghua Huang}, \bibinfo{person}{Mengyang Feng},
  \bibinfo{person}{Ningyi Zhang}, \bibinfo{person}{Pandeng Li},
  \bibinfo{person}{Pingyu Wu}, \bibinfo{person}{Ruihang Chu},
  \bibinfo{person}{Ruili Feng}, \bibinfo{person}{Shiwei Zhang},
  \bibinfo{person}{Siyang Sun}, \bibinfo{person}{Tao Fang},
  \bibinfo{person}{Tianxing Wang}, \bibinfo{person}{Tianyi Gui},
  \bibinfo{person}{Tingyu Weng}, \bibinfo{person}{Tong Shen},
  \bibinfo{person}{Wei Lin}, \bibinfo{person}{Wei Wang}, \bibinfo{person}{Wei
  Wang}, \bibinfo{person}{Wenmeng Zhou}, \bibinfo{person}{Wente Wang},
  \bibinfo{person}{Wenting Shen}, \bibinfo{person}{Wenyuan Yu},
  \bibinfo{person}{Xianzhong Shi}, \bibinfo{person}{Xiaoming Huang},
  \bibinfo{person}{Xin Xu}, \bibinfo{person}{Yan Kou}, \bibinfo{person}{Yangyu
  Lv}, \bibinfo{person}{Yifei Li}, \bibinfo{person}{Yijing Liu},
  \bibinfo{person}{Yiming Wang}, \bibinfo{person}{Yingya Zhang},
  \bibinfo{person}{Yitong Huang}, \bibinfo{person}{Yong Li},
  \bibinfo{person}{You Wu}, \bibinfo{person}{Yu Liu}, \bibinfo{person}{Yulin
  Pan}, \bibinfo{person}{Yun Zheng}, \bibinfo{person}{Yuntao Hong},
  \bibinfo{person}{Yupeng Shi}, \bibinfo{person}{Yutong Feng},
  \bibinfo{person}{Zeyinzi Jiang}, \bibinfo{person}{Zhen Han},
  \bibinfo{person}{Zhi-Fan Wu}, {and} \bibinfo{person}{Ziyu Liu}.}
  \bibinfo{year}{2025}\natexlab{}.
\newblock \bibinfo{title}{Wan: Open and Advanced Large-Scale Video Generative
  Models}.
\newblock
\showeprint[arxiv]{2503.20314}~[cs. CV]
\urldef\tempurl%
\url{https://arxiv.org/abs/2503.20314}
\showURL{%
\tempurl}


\bibitem[Wang et~al\mbox{.}(2016)]%
        {wang2016mcl}
\bibfield{author}{\bibinfo{person}{Haiqiang Wang}, \bibinfo{person}{Weihao
  Gan}, \bibinfo{person}{Sudeng Hu}, \bibinfo{person}{Joe~Yuchieh Lin},
  \bibinfo{person}{Lina Jin}, \bibinfo{person}{Longguang Song},
  \bibinfo{person}{Ping Wang}, \bibinfo{person}{Ioannis Katsavounidis},
  \bibinfo{person}{Anne Aaron}, {and} \bibinfo{person}{C-C~Jay Kuo}.}
  \bibinfo{year}{2016}\natexlab{}.
\newblock \showarticletitle{MCL-JCV: a JND-based H. 264/AVC video quality
  assessment dataset}. In \bibinfo{booktitle}{\emph{2016 IEEE international
  conference on image processing (ICIP)}}. IEEE, \bibinfo{pages}{1509--1513}.
\newblock


\bibitem[Xia et~al\mbox{.}(2024)]%
        {xia2024DiffPC}
\bibfield{author}{\bibinfo{person}{Yichong Xia}, \bibinfo{person}{Yimin Zhou},
  \bibinfo{person}{Jinpeng Wang}, \bibinfo{person}{Baoyi An},
  \bibinfo{person}{Haoqian Wang}, \bibinfo{person}{Yaowei Wang}, {and}
  \bibinfo{person}{Bin Chen}.} \bibinfo{year}{2024}\natexlab{}.
\newblock \showarticletitle{{{DiffPC}}: {{Diffusion-based High Perceptual
  Fidelity Image Compression}} with {{Semantic Refinement}}}. In
  \bibinfo{booktitle}{\emph{The {{Thirteenth International Conference}} on
  {{Learning Representations}}}}.
\newblock


\bibitem[Yang and Mandt(2023)]%
        {yang2023Lossy}
\bibfield{author}{\bibinfo{person}{Ruihan Yang} {and} \bibinfo{person}{Stephan
  Mandt}.} \bibinfo{year}{2023}\natexlab{}.
\newblock \showarticletitle{Lossy {{Image Compression}} with {{Conditional
  Diffusion Models}}}.
\newblock \bibinfo{journal}{\emph{Advances in Neural Information Processing
  Systems}}  \bibinfo{volume}{36} (\bibinfo{date}{Dec.} \bibinfo{year}{2023}),
  \bibinfo{pages}{64971--64995}.
\newblock


\bibitem[Yang et~al\mbox{.}(2024)]%
        {yangCogVideoXTexttoVideoDiffusion2024}
\bibfield{author}{\bibinfo{person}{Zhuoyi Yang}, \bibinfo{person}{Jiayan Teng},
  \bibinfo{person}{Wendi Zheng}, \bibinfo{person}{Ming Ding},
  \bibinfo{person}{Shiyu Huang}, \bibinfo{person}{Jiazheng Xu},
  \bibinfo{person}{Yuanming Yang}, \bibinfo{person}{Wenyi Hong},
  \bibinfo{person}{Xiaohan Zhang}, \bibinfo{person}{Guanyu Feng},
  \bibinfo{person}{Da Yin}, \bibinfo{person}{Xiaotao Gu},
  \bibinfo{person}{Yuxuan Zhang}, \bibinfo{person}{Weihan Wang},
  \bibinfo{person}{Yean Cheng}, \bibinfo{person}{Ting Liu},
  \bibinfo{person}{Bin Xu}, \bibinfo{person}{Yuxiao Dong}, {and}
  \bibinfo{person}{Jie Tang}.} \bibinfo{year}{2024}\natexlab{}.
\newblock \bibinfo{title}{{{CogVideoX}}: {{Text-to-Video Diffusion Models}}
  with {{An Expert Transformer}}}.
\newblock
\showeprint[arxiv]{2408.06072}~[cs]
\href{https://doi.org/10.48550/arXiv.2408.06072}{doi:\nolinkurl{10.48550/arXiv.2408.06072}}


\bibitem[Yin et~al\mbox{.}(2025)]%
        {yinSlowBidirectionalFast2025}
\bibfield{author}{\bibinfo{person}{Tianwei Yin}, \bibinfo{person}{Qiang Zhang},
  \bibinfo{person}{Richard Zhang}, \bibinfo{person}{William~T. Freeman},
  \bibinfo{person}{Fredo Durand}, \bibinfo{person}{Eli Shechtman}, {and}
  \bibinfo{person}{Xun Huang}.} \bibinfo{year}{2025}\natexlab{}.
\newblock \bibinfo{title}{From {{Slow Bidirectional}} to {{Fast Autoregressive
  Video Diffusion Models}}}.
\newblock
\showeprint{2412.07772}~[cs]
\href{https://doi.org/10.48550/arXiv.2412.07772}{doi:\nolinkurl{10.48550/arXiv.2412.07772}}


\bibitem[Zhang et~al\mbox{.}(2026)]%
        {zhang2026sageattention3microscalingfp4attention}
\bibfield{author}{\bibinfo{person}{Jintao Zhang}, \bibinfo{person}{Jia Wei},
  \bibinfo{person}{Pengle Zhang}, \bibinfo{person}{Xiaoming Xu},
  \bibinfo{person}{Haofeng Huang}, \bibinfo{person}{Haoxu Wang},
  \bibinfo{person}{Kai Jiang}, \bibinfo{person}{Jianfei Chen}, {and}
  \bibinfo{person}{Jun Zhu}.} \bibinfo{year}{2026}\natexlab{}.
\newblock \bibinfo{title}{SageAttention3: Microscaling FP4 Attention for
  Inference and An Exploration of 8-Bit Training}.
\newblock
\showeprint[arxiv]{2505.11594}~[cs.LG]
\urldef\tempurl%
\url{https://arxiv.org/abs/2505.11594}
\showURL{%
\tempurl}


\bibitem[Zhang et~al\mbox{.}(2025b)]%
        {turbodiffusion}
\bibfield{author}{\bibinfo{person}{Jintao Zhang}, \bibinfo{person}{Kaiwen
  Zheng}, \bibinfo{person}{Kai Jiang}, \bibinfo{person}{Haoxu Wang},
  \bibinfo{person}{Ion Stoica}, \bibinfo{person}{Joseph~E. Gonzalez},
  \bibinfo{person}{Jianfei Chen}, {and} \bibinfo{person}{Jun Zhu}.}
  \bibinfo{year}{2025}\natexlab{b}.
\newblock \bibinfo{title}{TurboDiffusion: Accelerating Video Diffusion Models
  by 100-200 Times}.
\newblock
\showeprint[arxiv]{2512.16093}~[cs.CV]
\urldef\tempurl%
\url{https://arxiv.org/abs/2512.16093}
\showURL{%
\tempurl}


\bibitem[Zhang et~al\mbox{.}(2025a)]%
        {zhang2025vsafastervideodiffusion}
\bibfield{author}{\bibinfo{person}{Peiyuan Zhang}, \bibinfo{person}{Yongqi
  Chen}, \bibinfo{person}{Haofeng Huang}, \bibinfo{person}{Will Lin},
  \bibinfo{person}{Zhengzhong Liu}, \bibinfo{person}{Ion Stoica},
  \bibinfo{person}{Eric Xing}, {and} \bibinfo{person}{Hao Zhang}.}
  \bibinfo{year}{2025}\natexlab{a}.
\newblock \bibinfo{title}{VSA: Faster Video Diffusion with Trainable Sparse
  Attention}.
\newblock
\showeprint[arxiv]{2505.13389}~[cs.CV]
\urldef\tempurl%
\url{https://arxiv.org/abs/2505.13389}
\showURL{%
\tempurl}


\bibitem[Zhang et~al\mbox{.}(2018)]%
        {zhang2018perceptual}
\bibfield{author}{\bibinfo{person}{Richard Zhang}, \bibinfo{person}{Phillip
  Isola}, \bibinfo{person}{Alexei~A Efros}, \bibinfo{person}{Eli Shechtman},
  {and} \bibinfo{person}{Oliver Wang}.} \bibinfo{year}{2018}\natexlab{}.
\newblock \showarticletitle{The Unreasonable Effectiveness of Deep Features as
  a Perceptual Metric}. In \bibinfo{booktitle}{\emph{CVPR}}.
\newblock


\bibitem[Zhao et~al\mbox{.}(2023a)]%
        {zhao2023dnervmodelinginherentdynamics}
\bibfield{author}{\bibinfo{person}{Qi Zhao}, \bibinfo{person}{M.~Salman Asif},
  {and} \bibinfo{person}{Zhan Ma}.} \bibinfo{year}{2023}\natexlab{a}.
\newblock \bibinfo{title}{DNeRV: Modeling Inherent Dynamics via Difference
  Neural Representation for Videos}.
\newblock
\showeprint[arxiv]{2304.06544}~[cs.CV]
\urldef\tempurl%
\url{https://arxiv.org/abs/2304.06544}
\showURL{%
\tempurl}


\bibitem[Zhao et~al\mbox{.}(2023b)]%
        {zhao2023unipcunifiedpredictorcorrectorframework}
\bibfield{author}{\bibinfo{person}{Wenliang Zhao}, \bibinfo{person}{Lujia Bai},
  \bibinfo{person}{Yongming Rao}, \bibinfo{person}{Jie Zhou}, {and}
  \bibinfo{person}{Jiwen Lu}.} \bibinfo{year}{2023}\natexlab{b}.
\newblock \bibinfo{title}{UniPC: A Unified Predictor-Corrector Framework for
  Fast Sampling of Diffusion Models}.
\newblock
\showeprint[arxiv]{2302.04867}~[cs.LG]
\urldef\tempurl%
\url{https://arxiv.org/abs/2302.04867}
\showURL{%
\tempurl}


\end{thebibliography}

\end{document}